\documentclass[fleqn,usenatbib]{mnras}

\usepackage{newtxtext,newtxmath}

\usepackage[T1]{fontenc}
\usepackage{subfig}

\DeclareRobustCommand{\VAN}[3]{#2}
\let\VANthebibliography\thebibliography
\def\thebibliography{\DeclareRobustCommand{\VAN}[3]{##3}\VANthebibliography}

\usepackage{graphicx}	
\usepackage{amsmath}	
\usepackage{appendix}
\usepackage{verbatim}

\newcommand{\rtt}[1]{\textcolor{black}{#1}}
\newcommand{\rt}[1]{\textcolor{black}{#1}}

\title[Discovery of USPs]{Discovery of small ultra-short-period planets orbiting \textit{Kepler} KG dwarfs with GPU phase folding and deep learning
}

\author[K. Wang et al.]{
Kaitlyn Wang$^{1,2}\thanks{E-mail:kaitwang@stanford.edu},$
Jian Ge$^{3}\thanks{E-mail:jge@shao.ac.cn},$
Kevin Willis$^{1},$
Kevin Wang$^{4},$
Yinan Zhao$^{5},$
Quanquan Hu$^{3,6}$
\\
$^{1}$Science Talent Training Center, Gainesville, FL, 32606 USA\\
$^{2}$Stanford University, 450 Jane Stanford Way, Stanford, CA, 94305 USA\\
$^{3}$Shanghai Astronomical Observatory, Shanghai 200030, China\\
$^{4}$Princeton University, PO Box 430 Princeton, NJ 08544, USA\\
$^{5}$Department of Astronomy, University of Geneva, Switzerland\\
$^{6}$University of Chinese Academy of Sciences, No. 19A Yuquan Road, Beijing 100049, China
}

\date{Accepted XXX. Received YYY; in original form ZZZ}

\pubyear{2023}

\begin{document}
\label{firstpage}
\pagerange{\pageref{firstpage}--\pageref{lastpage}}
\maketitle


\begin{abstract}
Of over 5,000 exoplanets identified so far, only a few hundred possess sub-Earth radii. The formation processes of these sub-Earths remain elusive, and acquiring additional samples is essential for investigating this unique population. In our study, we employ the GPFC method, a novel GPU Phase Folding algorithm combined with a Convolutional Neural Network, on \textit{Kepler} photometry data. This method enhances the transit search speed significantly over the traditional Box-fitting Least Squares method, allowing a complete search of the known \textit{Kepler} KOI data within days using a commercial GPU card. To date, we have identified five new ultra-short-period planets (USPs): Kepler-158d, Kepler-963c, Kepler-879c, Kepler-1489c, and KOI-4978.02. Kepler-879c with a radius of 0.4 R$_\oplus$ completes its orbit around a G dwarf in 0.646716 days.  Kepler-158d with a radius of 0.43 R$_\oplus$ orbits a K dwarf star every 0.645088 days.  Kepler-1489c with a radius of 0.51 R$_\oplus$ orbits a G dwarf in 0.680741 days. Kepler-963c with a radius of 0.6 R$_\oplus$ revolves around a G dwarf in 0.919783 days, and KOI-4978.02 with a radius of 0.7 R$_\oplus$ circles a G dwarf in 0.941967 days. Among our findings, Kepler-879c, Kepler-158d and Kepler-963c rank as the first, the third, the fourth smallest USPs identified to date. Notably, Kepler-158d stands as the smallest USP found orbiting K dwarfs while Kepler-963c, Kepler-879c, Kepler-1489c, and KOI-4978.02 are the smallest USPs found orbiting G dwarfs.  Kepler-879c, Kepler-158d, Kepler-1489c, and \rtt{KOI-4978.02} are among the smallest planets that are closest to their host stars, with orbits within 5 stellar radii. In addition, these discoveries highlight GPFC’s promising capability in identifying small, new transiting exoplanets within photometry data from \textit{Kepler}, \textit{TESS}, and upcoming space transit missions, \textit{PLATO} and \textit{ET}.

\end{abstract}




\section{Introduction}

Contrary to prevailing solar system-based planet formation theories, which do not predict planets with orbits significantly \rt{closer} than Mercury's, Ultra-short-period Planets (USPs) represent a captivating group of exoplanets with orbital periods of less than one day. These USPs may offer vital insights into the early evolution of planetary systems and the dynamics of star-planet interactions, including tidal forces and atmospheric erosion. The most extreme USPs, e.g. KOI-1843.03 (K2-137b), have orbital periods of just 4 hours, bordering on the edge of tidal disruption \citep{Rappaport_2013, Smith_2017}. Found in approximately 0.5\% of Sun-like stars, USPs typically have radii smaller than 2 R$_\oplus$ or, in the case of ultra-hot Jupiters, greater than 10 R$_\oplus$ \citep{Sanchis_Ojeda_2014}. Understanding how these planets achieve such short orbital periods remains one of the longest-standing unresolved problems in the field.

To date, three leading theories have been proposed to explain the formation of USPs: in situ formation, migration driven by stellar tides, and tidal migration coupled with multi-body interactions \citep{Adams_2021}. The in situ formation theory, which posits that these planets formed near their present orbits, is deemed less plausible due to the extreme temperatures these close-in orbits would encounter \citep{Boss_1998}. Moreover, given that host stars' radii were significantly larger during their pre-main sequence phase compared to now, USPs in closer proximity would likely have been engulfed by their host stars \citep{Palla_Stahl_1991, DAntona_Mazzitelli_1994}. Meanwhile, the role of tidal migration in the genesis of these planets remains an active area of investigation \citep{Dawson_2018}. Given that USPs are often observed with outer accompanying planets in longer orbits, it is hypothesized that USP origins involve interactions among the sibling planets, which reposition USPs into their current positions, closer to their host stars, possibly in orbits previously occupied by the stars themselves. \rt{\citet{Lee_2017} stated that planets can \rtt{migrate} from disk edges to ultra-short periods through asynchronous equilibrium tides raised on their stars and that tidal migration contributes to the wider orbital spacing of USPs compared to their longer-period counterparts.}

Regardless of its origin, once a planet settles into a short-period orbit, the host star’s influence may engulf it in a few billion years. For volatile-rich planets, high atmospheric temperatures and intense stellar irradiation they endure can lead to atmospheric loss \citep{Owen_2019}. Additionally, strong tidal forces can cause orbital decay, pushing these planets into a state of unstable Roche lobe overflow. Such destabilization can culminate in the complete disruption of their atmospheres after just a few orbits \citep{Jia_2016}.

The main limitation in studying the enigmatic USPs has been the scarcity of USPs discovered so far. To date, there are only 61 USPs confirmed by the \textit{Kepler} survey and 147 in the \href{https://exoplanetarchive.ipac.caltech.edu/}{NASA Exoplanet Archive}. Assessing the relative abundance and properties of USPs, including variations in orbital periods, planetary mass, the spectral type of their host stars, and the configurations of any associated multiplanet systems is vital for testing theoretical models. However, the small sample size of known USPs means that current estimates of their occurrence rates are subject to large error margins. Therefore, expanding the USP data set is imperative for gaining a deeper understanding of this group.

In this paper, we present the discovery and detailed analysis of five new USPs—Kepler-158d, Kepler-963c, Kepler-879c, Kepler-1489c, and KOI-4978.02. \rt{Upon eventual official confirmation, Kepler-879c, Kepler-158d, Kepler-1489c, and \rtt{KOI-4978.02} are among the smallest planets that are closest to their host stars, with orbits within 5 stellar radii. Additionally, Kepler-879c, Kepler-158d and Kepler-963c rank as the first, the third, the fourth smallest USPs detected to date. }

Our paper is organized as follows: Section 2 provides a brief overview of the Graphics Processing Unit (GPU) Phase Folding and Convolutional Neural Network (GPFC) transit search method that led to these discoveries. Section 3 details our systematic approach to planet discovery, including the pre-processing of \textit{Kepler} data, phase folding, and transit detection. In Section 4, we introduce new USPs, describing both the vetting and fitting processes used for validation and parameter determination. Section 5 delves into an in-depth discussion of the USPs' characteristics. Finally, Section 6 draws conclusions and future  potential applications of our GPFC method.

\section{Overview of the GPFC method}

The GPFC method, as detailed in \citet{kwang2023gpu} and applied to the \textit{Kepler} survey dataset, has produced the USP discoveries presented in this paper. GPFC combines a novel GPU phase folding algorithm with a Convolutional Neural Network (CNN) to enhance transit detection. A key advantage of GPFC is its proficiency in detecting low Signal-to-Noise Ratio (SNR) planetary signals by processing raw light curves directly from the \textit{Kepler} survey, without reliance on pre-identified signal catalogs like the \textit{Kepler} Threshold Crossing Event (TCE) catalog.

The GPFC method comprises four main stages specifically designed to identify transiting exoplanet candidates from continuous photometric time-series surveys. It begins with preprocessing the \textit{Kepler} Pre-search Data Conditioning Simple Aperture Photometry (PDCSAP) light curves, which have been treated to mitigate instrumental variations and capture the intrinsic stellar brightness over time. Our preprocessing step involves removing outliers and correcting for long-term stellar variability. Following this, the light curves are phase-folded at various trial periods, which are evenly distributed across the USP search range of [0.2 1] days, utilizing the GPU phase folding technique. The resulting folded light curves are then scaled and fed into the CNN module, which assesses the presence of potential exoplanet transits at these trial periods. Light curves that receive high scores from the CNN module are further examined during the vetting stage.

\rt{Our GPU-based phase folding algorithm was executed on an Nvidia GeForce GPU card, which supports a grid dimension configuration of up to ($2^{31}-1, 65535, 65535$) for parallel processing. At 100,000 search frequencies, the GPFC method operated 15 times faster than the latest \href{docs.astropy.org}{AstroPy implementation} of the predominant Box-fitting Least Squares (BLS) method \citep{Kov_cs_2002}, which has been enhanced with {\sc Cython} acceleration. This speed increase \rtt{would reduce} the search time for all of the $1-2 \times 10^5$ targets in the \textit{Kepler} Archive from $3.4-6.7$ months to $7-14$ days.}

\rt{Meanwhile, GPFC demonstrates higher detection accuracy, with a 7\% higher Area Under the Receiver Operating Characteristic (ROC) Curve (AUC) compared to BLS. At SNR 7, GPFC also shows a 7\% higher true positive rate for the same false positive rate (FPR) of detection, along with higher precision at the same recall rate \citep{kwang2023gpu}.}

In a blind search of \textit{Kepler} light curves, GPFC successfully recovers 100\% of confirmed USP planets in \textit{Kepler} and the 106 USPs listed in \citet{Sanchis_Ojeda_2014}.

\begin{figure}
    \centering
    \includegraphics[scale = 0.32] {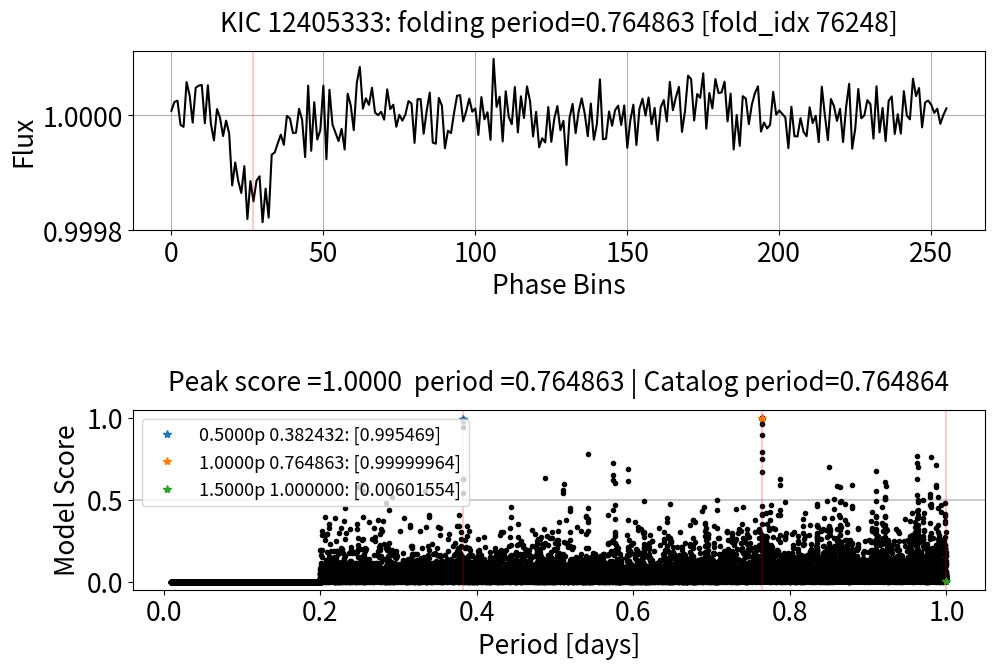}
    \caption{Kepler-1409b completes an orbit around its host star, KIC 12405333, every 0.764863 days. This transit period was accurately identified using the GPFC method. The top panel illustrates a clear transit signal corresponding to the 0.764863-day period. The bottom panel shows a peak CNN score of 1.0000 aligning with this period, with another notable score of 0.9955 observed at the harmonic half-period of 0.382432 days.} 
    \label{fig:real_usp_yt_illustrate}
\end{figure}

\begin{figure} 
    \centering
    \includegraphics[scale = 0.32] {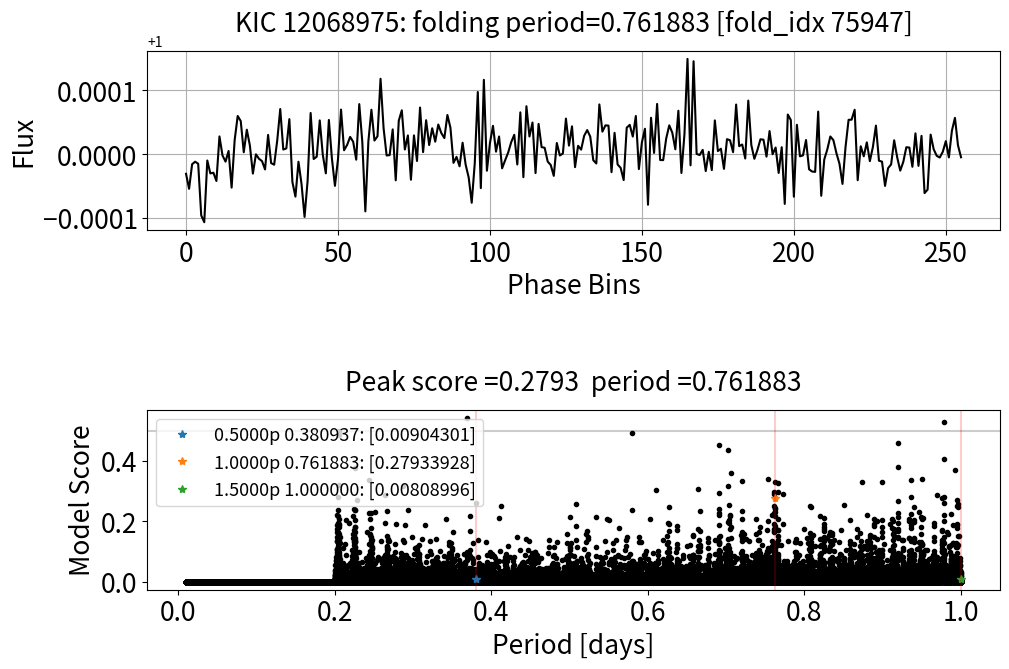}
    \caption{Depiction of the folded light curve for KIC 12068975, which has no known USP transits and serves as an example where no transit signal is detected within our search range of [0.2,1] days. The most prominent CNN score recorded is 0.2793 at a period of 0.761883 days, which is classified as a non-detection by the GPFC system.}
    \label{fig:no_transit_flux_cnn}
\end{figure}

We demonstrate the efficacy of the GPFC method using two \textit{Kepler} objects from the \textit{Kepler} Objects of Interest (KOI) catalog. First, KIC 12405333: a star hosting a confirmed USP planet with an orbital period of 0.764863 days. As shown in Figure \ref{fig:real_usp_yt_illustrate}, the folded light curve generated by our method matches the orbital period documented in the KOI catalog, yielding a peak CNN score of 1.0000. The folded light curve prominently displays the transit-induced flux dip. In contrast, KIC 12068975 is taken as an example which does not have any known planet transits within the range of [0.2,1] days. By removing all known transit signals from the light curve, we ensure no harmonics will interfere within our USP search range. Figure \ref{fig:no_transit_flux_cnn} presents the results: a peak CNN score of a mere 0.2793 at the 0.761883-day fold period with an absence of any discernible transit signal in the folded light curve.

\section{Transit Search Process}
\rt{In this section, we describe the steps our GPFC system takes to identify USP exoplanets within the \textit{Kepler} survey dataset, which include preprocessing raw \textit{Kepler} light curves, phase folding, and employing a CNN to assess the potential presence of exoplanet transits. Throughout the stages of our GPFC system, we extensively utilized the Lightkurve package \citep{Lightkurve_pkg}, particularly the foundational astronomy libraries of \href{https://astropy.org/}{Astropy} and \href{https://astroquery.readthedocs.io/}{Astroquery}.}

\subsection{\textit{Kepler} Dataset Selection}
Initially, light curves from the Q1–Q17 \textit{Kepler} Data Release 25 (DR25) were retrieved, sourced from the \href{https://archive.stsci.edu/}{Mikulski Archive for Space Telescopes}. Generated by the \textit{Kepler} Science Processing Pipeline \citep{Jenkins_2010}, these light curves contain integrated flux measurements taken at 1766-second intervals, spanning approximately four years and encompassing 30,000 to 70,000 epochs. Additionally, the \textit{Kepler} Objects of Interest (KOIs) dataset, as recorded in the NASA Exoplanet Archive as of June 2023, was also downloaded. This archive contained 9564 dispositioned KOIs, including 2350 confirmed planets, 2366 candidates, and 4848 false positives. For our analysis, we utilized transit parameters such as period, epoch, and duration from the KOI catalog to assist in light curve processing and transit verification.

We filtered our light curve dataset using the KOI catalog, excluding stars hosting planets labeled as 'FALSE POSITIVE' in the Exoplanet Archive Disposition field. Our study was focused on stars with all planets designated as 'CONFIRMED' or 'CANDIDATE'. This approach ensured our dataset was mitigated from secondary eclipses caused by eclipsing binaries, which are not documented in the KOI catalog and could potentially introduce contamination. This filtering substantially reduced the risk of false positive detection in our study.

\subsection{Data Pre-processing}
For each retrieved light curve, we conducted pre-processing akin to the procedure outlined in \citet{Shallue_2018}. Initially, we generated a "non-transit" light curve by excluding all known transits (including confirmed, candidate and false positive dispositions) listed in the KOI catalog, preparing the light curve for potential new planet discoveries. The light curve was then detrended by iteratively fitting it with a B-spline, with 3-$\sigma$ outliers discarded progressively. Subsequently, the light curve was normalized by dividing it by the best-fit spline. This normalized light curve served as the input for our fast GPU phase-folding module.

\subsection{Fast GPU Phase Folding}
Our fast GPU phase-folding module conducts phase folding on non-transit light curves by iterating across 100,000 uniformly distributed trial periods, focusing on the [0.2 1] days range for USPs while remaining adaptable to broader ranges. Each light curve is folded and binned at these trial periods, resulting in 100,000 folded light curves with 256 bins each. The precision in phase folding is crucial, as it ensures the capture of subtle signals from shallow transits, minimizing the risk of missing potential detections due to small time offsets. The GPU's grid system allows for the simultaneous folding of the 100,000 trial periods, requiring approximately 5 seconds to fold a light curve with around 65,000 data points.

\subsection{Noise Normalization}
After obtaining the 100,000 folded light curves from the GPU Phase Folding module, we proceeded with a noise normalization process. This involved scaling the flux values to achieve a standard deviation of 1.0 while maintaining the mean of 1.0 to match the standard deviation and mean of the synthetic data used in neural network training. Ensuring uniformity in standard deviation across all light curves is crucial for the subsequent analysis by the Convolutional Neural Network module, optimizing them for more effective and consistent pattern recognition.

\subsection{Convolutional Neural Network}
Following noise normalization, we inputted the 100,000 folds into our CNN module, which was specifically trained to detect USP transit signals commonly observed in \textit{Kepler} light curves. Our training dataset comprised two million synthetic light curves: half included USP transit signals crafted to emulate the parameter distributions (such as period, duration, and radius ratio) observed in \textit{Kepler}'s confirmed USP transits, and the other half were devoid of any transit signals. Post-training, the best-performing model achieved an accuracy of 94.6\%. We utilized the Adam optimization algorithm \citep{Kingma_Adam_2014} to minimize the cross-entropy error function, with a learning rate set at $10^{-6}$ and a batch size of 32 across 90 epochs. Recognizing that the CNN might also pick up weaker peaks related to the harmonics of the transit period, we also performed a check to confirm that the CNN score identified had the highest score among its harmonics.

\section{Results}

Upon completing the validation process for transit events flagged with high confidence by the CNN module, we identified five robust candidates, as listed in Table~\ref{tab:robust_candidates}. The selection of these candidates hinged on two key criteria: firstly, they each obtained high scores from the CNN module, surpassing a defined threshold of 0.5. Secondly, these candidates' validity as transiting exoplanets was examined through a rigorous vetting process outlined in the subsequent section.

\begin{table}
    \centering
    \caption{Candidate List: KOI, KIC, detected orbital period, CNN model score, transit SNR, and associated False Alarm Probability (FAP).}
    \begin{tabular}{ l | l | c | c | c | c }
    \hline
    \hline
    \multicolumn{1}{c|}{\centering\textbf{KOI}} & \multicolumn{1}{c|}{\centering\textbf{KIC}} & \multicolumn{1}{p{0.8cm}|}{\centering\textbf{Period [days]}} & \multicolumn{1}{p{0.9cm}|}{\centering\textbf{CNN Score}} & \multicolumn{1}{p{0.8cm}|}{\centering\textbf{Transit SNR}} & \multicolumn{1}{p{0.7cm}|}{\centering\textbf{FAP [\%]}} \\
    \hline
    Kepler-963c & \phantom{0}8832512 & 0.919783 & 0.9309 & 7.3 & 0.118 \\
    Kepler-1489c & \phantom{0}8409295 & 0.680741 & 0.5599 & 6.2 & 0.463 \\
    Kepler-158d & \phantom{0}4633570 & 0.645088 & 0.8918 & 6.5 & 0.700 \\
    Kepler-879c & 12266636 & 0.646716 & 0.9650 & 6.3 & 0.792 \\
    KOI-4978.02 & \phantom{0}3428127 & 0.941967 & 0.8308 & 6.9 & 0.922 \\
    \hline
    \end{tabular}
    \label{tab:robust_candidates}
\end{table}

\begin{figure}
    \centering
    \includegraphics[scale = 0.32]{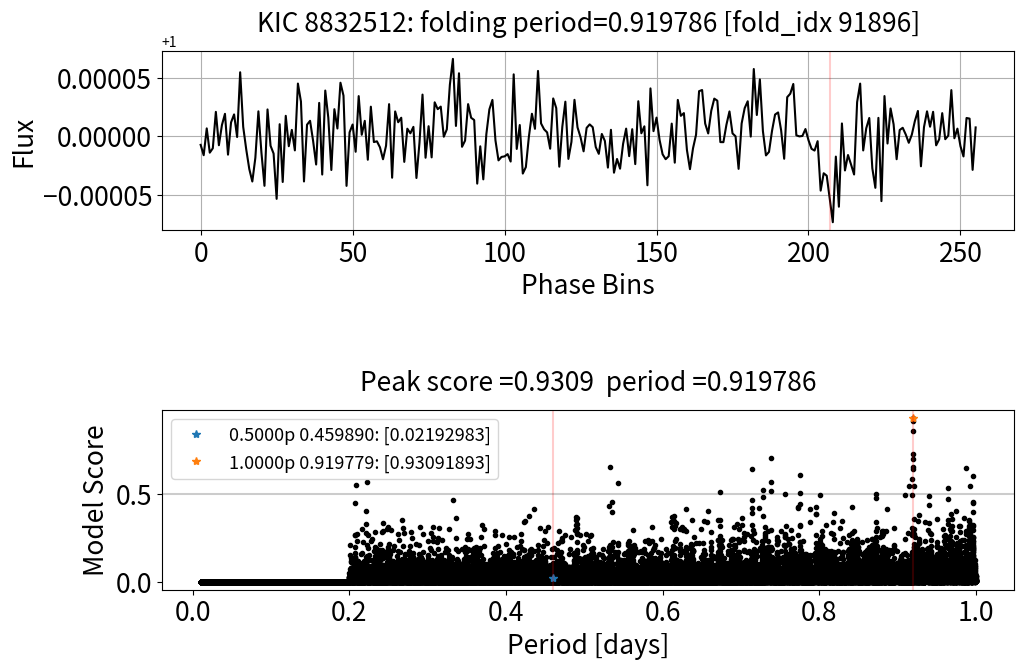}
    \caption{Folded light curve and CNN scores for candidate Kepler-963c orbiting host star KIC 8832512. A transit within the USP period range [0.2 1] days is detected at 0.919783 days, featuring a CNN score of 0.9309 and a transit SNR of 7.3. The detected transit is marked with a red line in the upper Flux plot. The CNN scores at the detected period and corresponding harmonics are marked in redlines in the lower Model Score plot.}
    \label{fig:8832512_K01821.b_k963c_flux_cnn}
\end{figure}

\begin{figure}
    \centering
    \includegraphics[scale = 0.32]{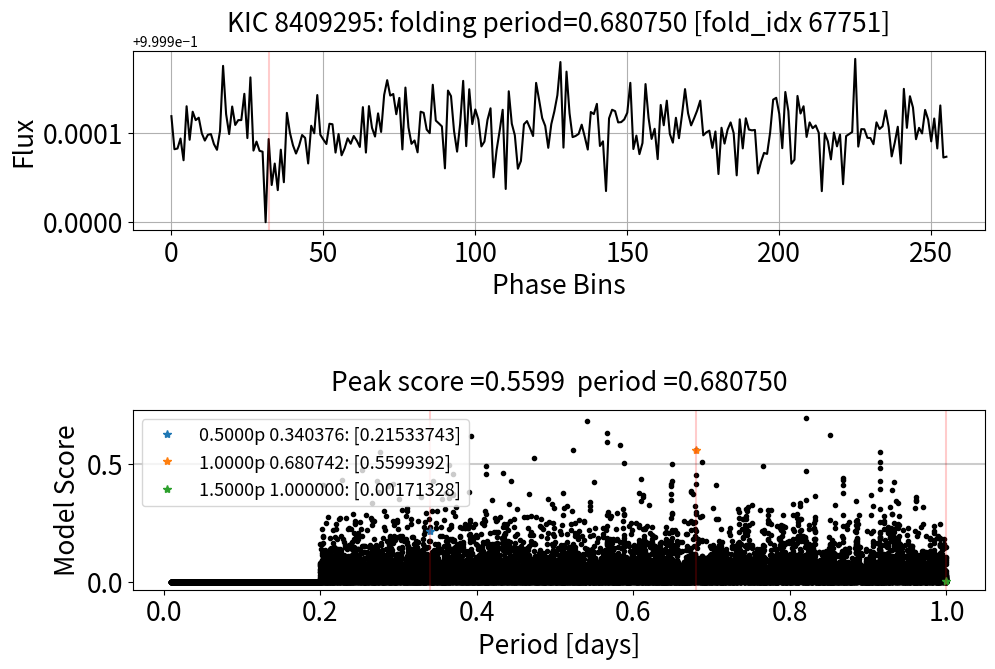}
    \caption{Folded light curve and CNN scores for Kepler-1489c orbiting KIC 8409295. A transit is detected within the USP period range [0.2 1] days at 0.680741 days, with a CNN score of 0.5599 and a transit SNR of 6.2. The detected transit is marked with a red line in the upper Flux plot. The CNN scores at the detected period and corresponding harmonics are marked in redlines in the lower Model Score plot.}
    \label{fig:8409295_K03404.b_k1489c_flux_cnn}
\end{figure}

\begin{figure}
    \centering
    \includegraphics[scale = 0.32]{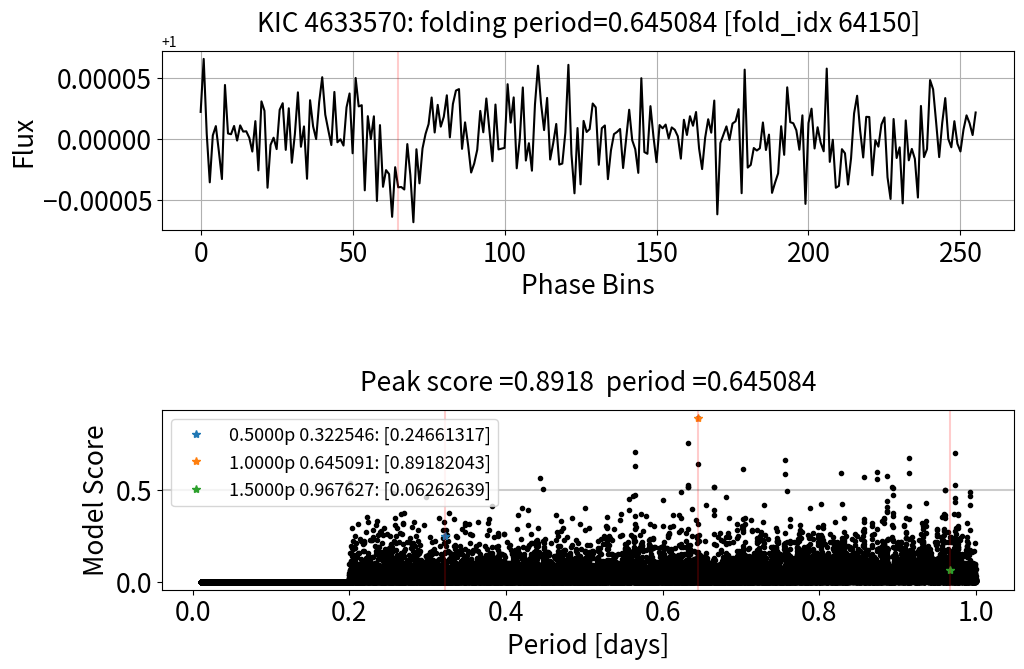}
    \caption{Folded light curve and CNN scores for Kepler-158d orbiting KIC 4633570. A transit is detected within the USP period range [0.2 1] days at 0.645088 days, with a CNN score of 0.8918 and a transit SNR of 6.5. The detected transit is marked with a red line in the upper Flux plot. The CNN scores at the detected period and corresponding harmonics are marked in redlines in the lower Model Score plot.}
    \label{fig:4633570_K00446.c_k158d_flux_cnn}
\end{figure}

\begin{figure}
    \centering
    \includegraphics[scale = 0.32]{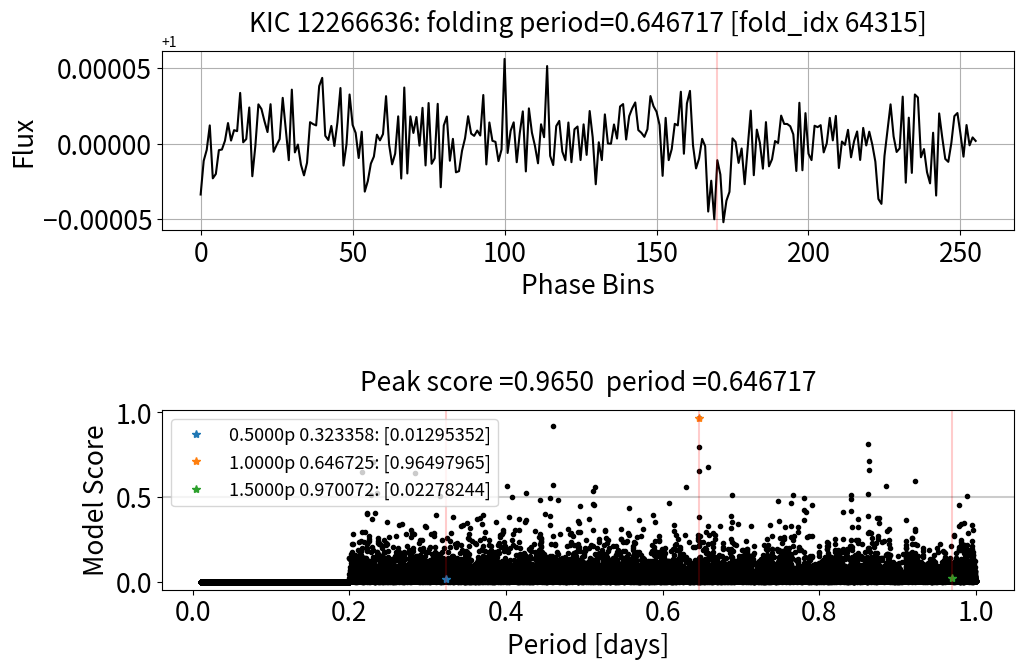}
    \caption{Folded light curve and CNN scores for candidate Kepler-879c orbiting host star KIC 12266636. A transit within the USP period range [0.2 1] days is detected at 0.646716 days, featuring a CNN score of 0.9650 and a transit SNR of 6.3. The detected transit is marked with a red line in the upper Flux plot. The CNN scores at the detected period and corresponding harmonics are marked in redlines in the lower Model Score plot.}
    \label{fig:12266636_k0152.03_k879d_flux_cnn}
\end{figure}

\begin{figure}
    \centering
    \includegraphics[scale = 0.32]{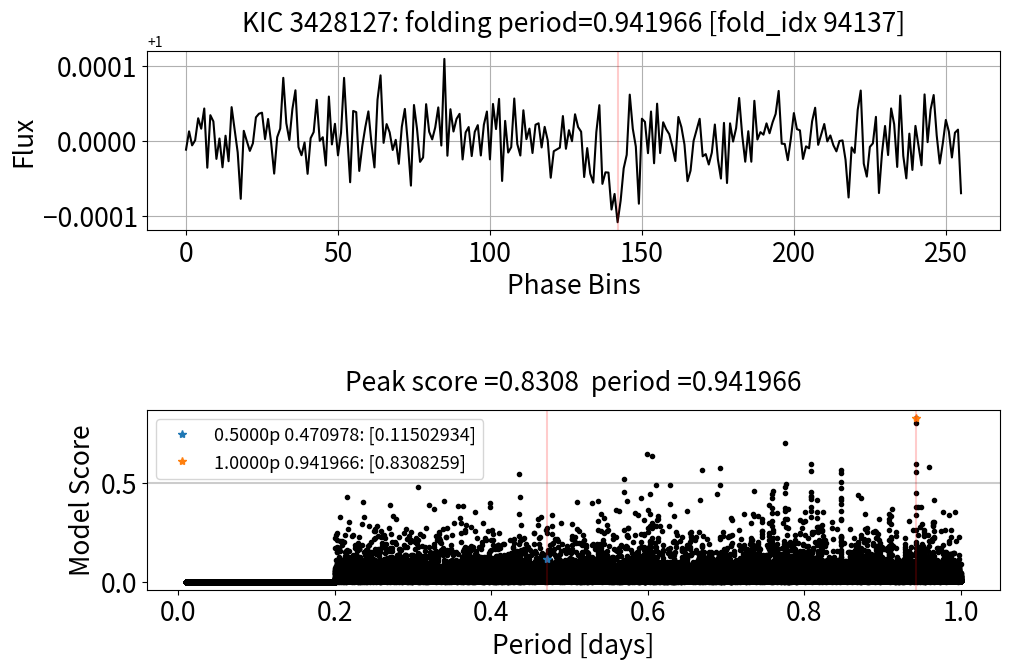}
    \caption{Folded light curve and CNN scores for candidate KOI-4978.02 orbiting host star KIC 3428127. A transit within the USP period range [0.2 1] days is detected at 0.941967 days, featuring a CNN score of 0.8308 and a transit SNR of 6.9. The detected transit is marked with a red line in the upper Flux plot. The CNN scores at the detected period and corresponding harmonics are marked in redlines in the lower Model Score plot.}
    \label{fig:3428127_K04978.b_n_flux_cnn}
\end{figure}

\subsection{Candidate Validation Process}
\label{sec:vetting}

In this study, we subjected potential exoplanet candidates to a comprehensive vetting process comprising ten tests, as detailed below. The tests are compiled from \href {https://exoplanetarchive.ipac.caltech.edu/docs/KeplerDV.html}{\textit{Kepler} Data Validation Documentation} in the NASA Exoplanet Archive, prior research \citep{Shallue_2018, Shahaf_2022, Adams_2021}, and methodologies developed by our team. As an illustrative example, we demonstrate the vetting results for candidate Kepler-963c throughout this section, with the same checks being performed for each candidate.

\begin{figure}
    \centering
    \begin{minipage}[c]{0.42\textwidth}
    \centering
    \subfloat{\includegraphics[width=\textwidth]{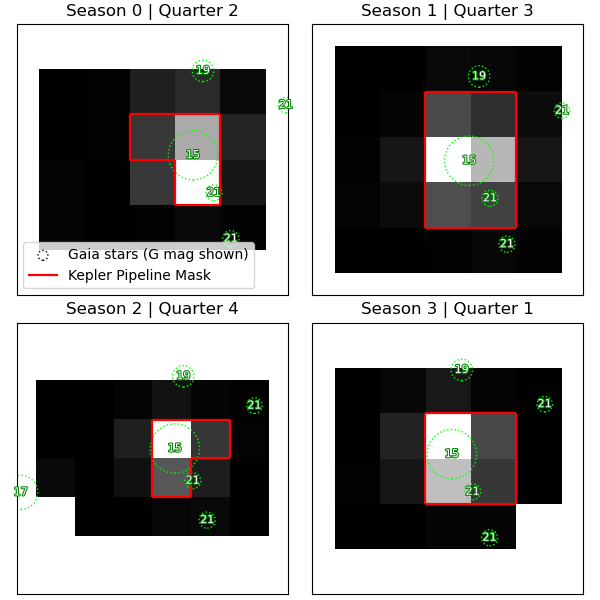}}\label{fig:8832512_seasonal_tpfs}
    \end{minipage}
    \vfill
    \begin{minipage}[c]{0.42\textwidth}
    \centering
    \subfloat{\includegraphics[width=\textwidth]{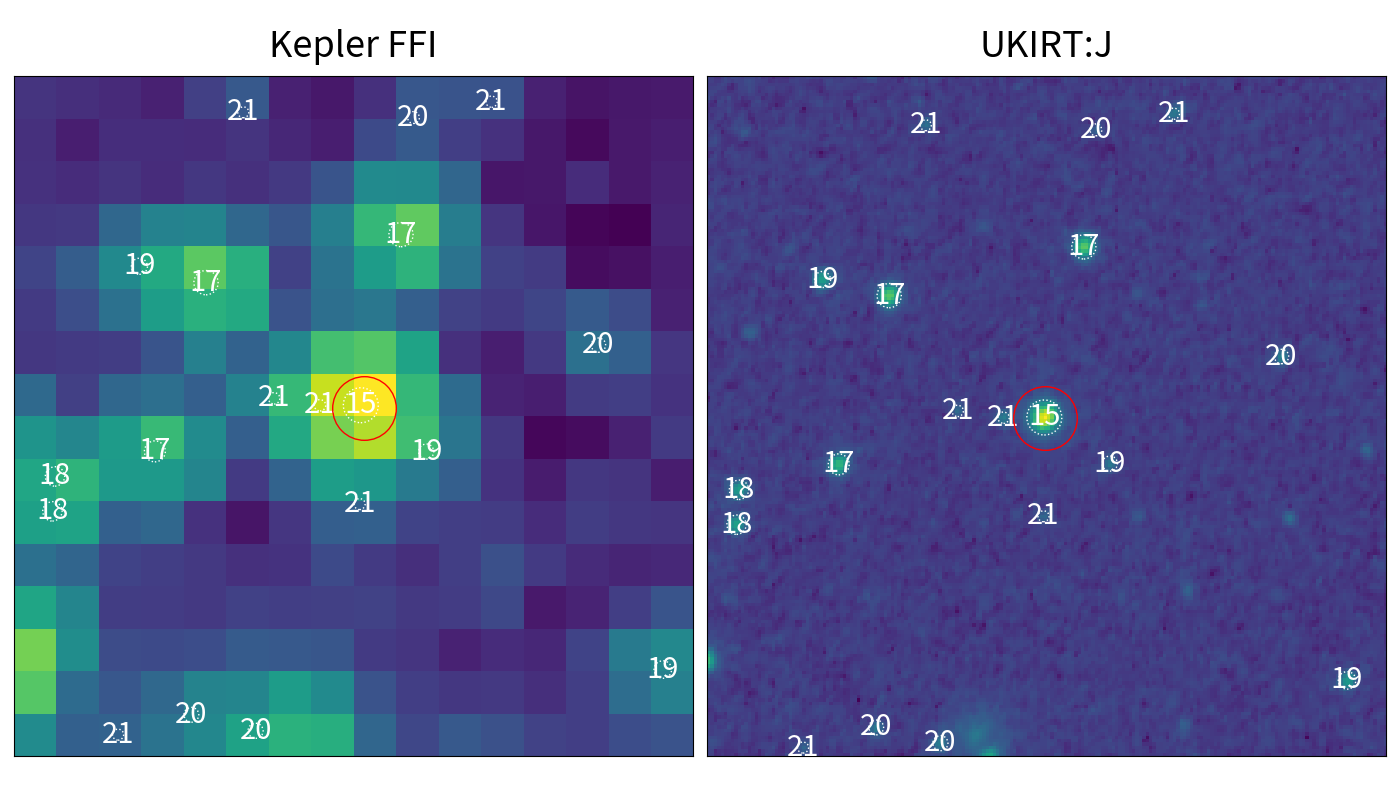}}\label{fig:8832512_ffi_ukirtJ}
    \end{minipage}
    \caption{The seasonal Target Pixel Files (TPF), Full-field Image (FFI) and UKIRT:J plots for candidate Kepler-963c. The upper four seasonal TPFs show the position and apparent magnitude value (Gaia) of the neighboring stars. Neighboring stars located within the target's photometric optimal aperture (red contour) exhibit significantly higher magnitudes, indicating low flux contamination. The FFI and UKIRT:J images below provide a broader view of the field, allowing for visual inspection without any apparent bright stars in these seeing-limited images.}
    \label{fig:8832512_tfps_ffi_ukirtj}
\end{figure}

\textbf{1. Neighbor Star Contamination Check}\newline
We first investigated the possibility of the transit signal arising from contamination by nearby stars \citep{Shallue_2018}. For this purpose, we acquired Target Pixel File (TPF) images from the \textit{Kepler} Community Follow-up Program (CFOP) and examined them for any contamination from neighboring stars within a 20-arcsecond radius. An initial assessment of contamination was conducted using seeing-limited images from UKIRT. Further examination involved detailed analysis of available high-contrast images using adaptive optics at the Lick Shane 3-m telescope (such as KIC 8832512). Figure \ref{fig:8832512_tfps_ffi_ukirtj} presents the seasonal TPFs, FFI, and UKIRT:J imagery of Kepler-963c.

\textbf{2. Secondary Eclipse Check}\newline
We examined the phase-folded light curves for signs of secondary eclipses, which could indicate that the transit signal originates from an eclipsing binary rather than a planet \citep{Shallue_2018}. After masking the USP transits, we searched for secondary eclipses using the BLS method. No secondary eclipses were observed for any of the host stars of our exoplanet candidates.

\textbf{3. BLS Peaks and Known Sibling Transit Harmonics Check}\newline
\rt{Subsequently, we utilized the BLS method to analyze the light curves and examine the peaks in the Signal Detection Efficiency (SDE) metrics of BLS. Using the same high-precision search frequency employed by GPFC, we conducted searches with GPU-BLS, a GPU-accelerated variant of the BLS method. We confirmed that GPU-BLS successfully identified our five USP candidates at the same periods detected by the GPFC method. \rtt{Note that these USP candidates were not included in the existing \textit{Kepler} TCE catalog, indicating that the catalog was generated with a search of lower frequency of trial periods, which yielded SNRs for small planets below the detection threshold.}}

For each of our candidates, we examined the BLS peaks at the detected transit period harmonics to ensure they did not overlap with the harmonics of the known transit periods of other sibling planets in the host planetary system. Similarly, we verified that these peaks do not coincide with the BLS SDE peaks at the harmonics of the host star's stellar rotation period.

\begin{figure*}
    \centering
    \includegraphics[scale = 0.45]{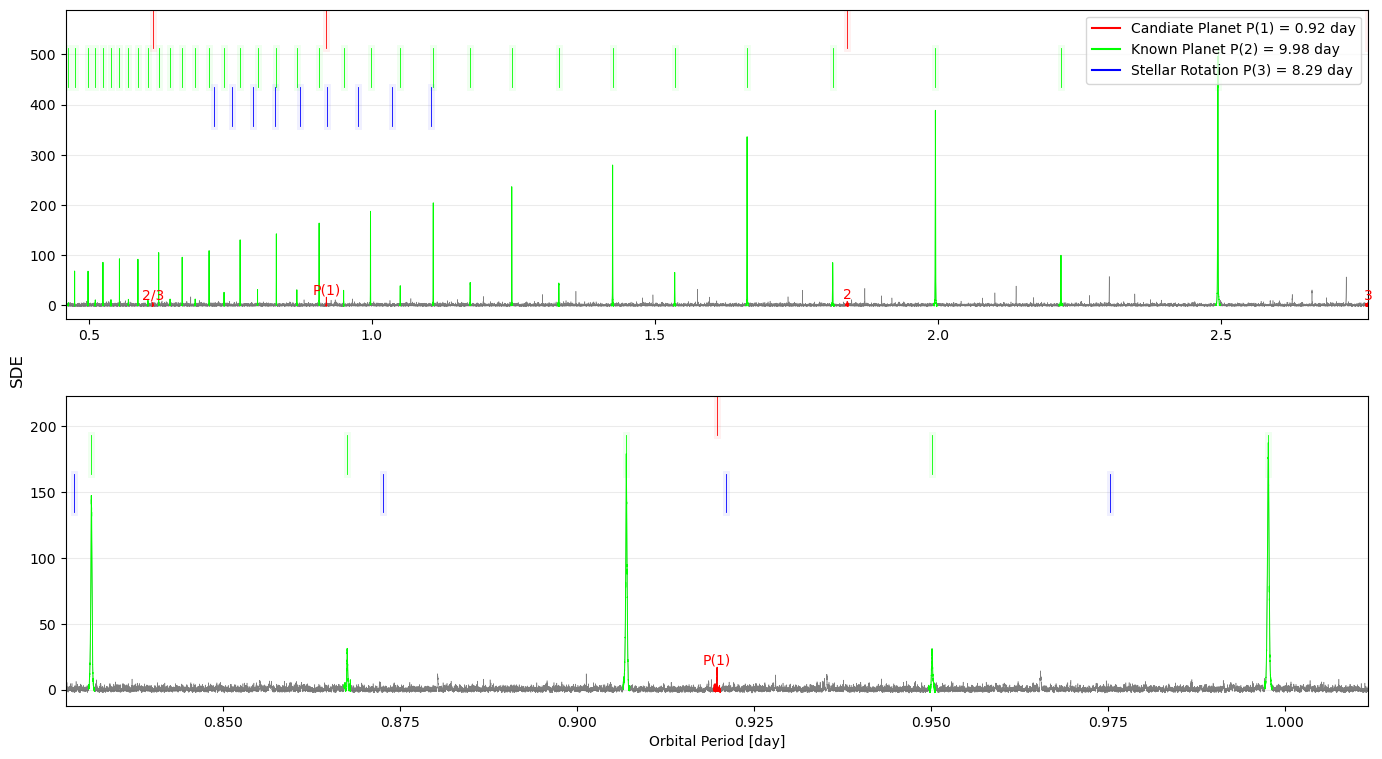}
    \caption{This figure illustrates the BLS Peaks Check and Known Sibling Transit Harmonics Check for candidate Kepler-963c. Harmonics of the orbital period for the host star's known orbiting planet are color-coded in green, while harmonics associated with the rotational period of the variable star are indicated in blue. Meanwhile, the harmonics of the orbital period of the new candidate are marked in red. This visualization confirms that the orbital period of the new candidate does not coincide with the periods of the already-known planet or the rotational period of the star.}
    \label{fig:8832512_vet_bls}
\end{figure*}

Figure \ref{fig:8832512_vet_bls} presents the BLS periodogram, with annotations marking the anticipated positions of both harmonics and fractional harmonics for our candidate transits, alongside those of transit signals already known to exist in the light curve. These annotations assist in discerning whether the observed transits of the candidates could stem from transits of other known planets.

\begin{figure}
    \centering
    \includegraphics[scale = 0.27]{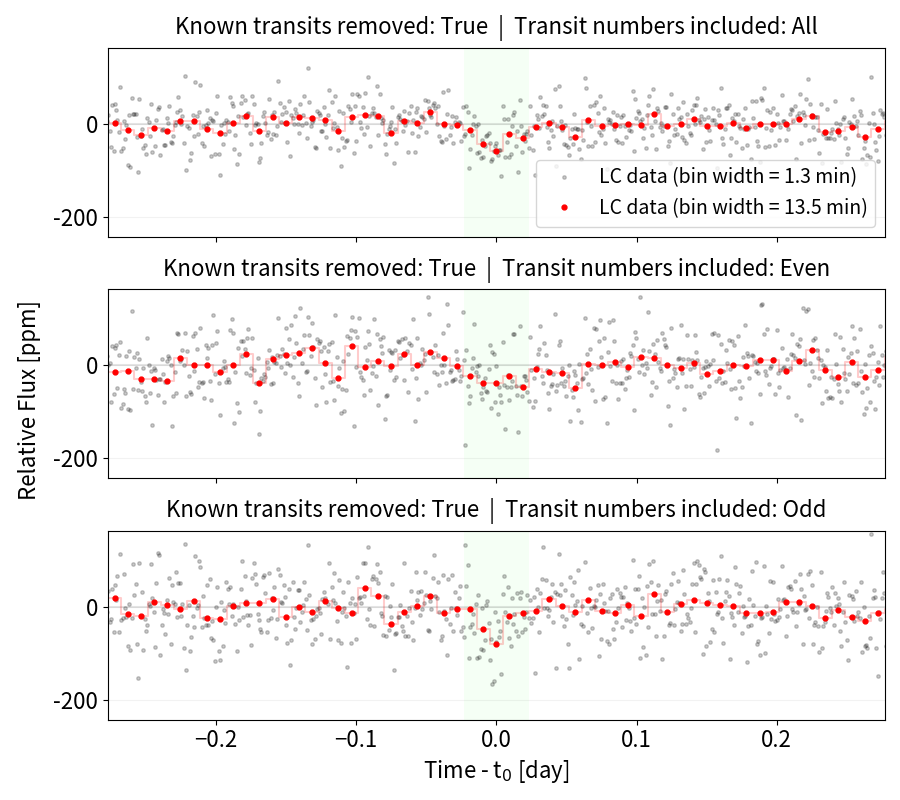}
    \caption{Even-odd Transit Check for candidate Kepler-963c. This test verifies that the transit signals of the new can didates remain consistent across both subsets of the light curve time sequence.}
    \label{fig:8832512_even_odd}
\end{figure}

\textbf{4. Even-odd Transits Check}\newline
Following this, we performed an even-odd transit analysis, wherein the data were folded and examined separately for even-numbered and odd-numbered transits, following the methodology outlined by \citet{Shallue_2018}. Our analysis confirmed consistent transit signals across both the even and odd sets for each candidate, with no significant discrepancies in transit depths, central transit time ($t_0$), or overall transit profiles. This consistency check is essential for eliminating potential false positives resulting from instrumental or other non-astronomical anomalies. Figure \ref{fig:8832512_even_odd} depicts the results of this even-odd transit analysis for candidate Kepler-963c.

\begin{figure}
    \centering
    \includegraphics[scale = 0.27]{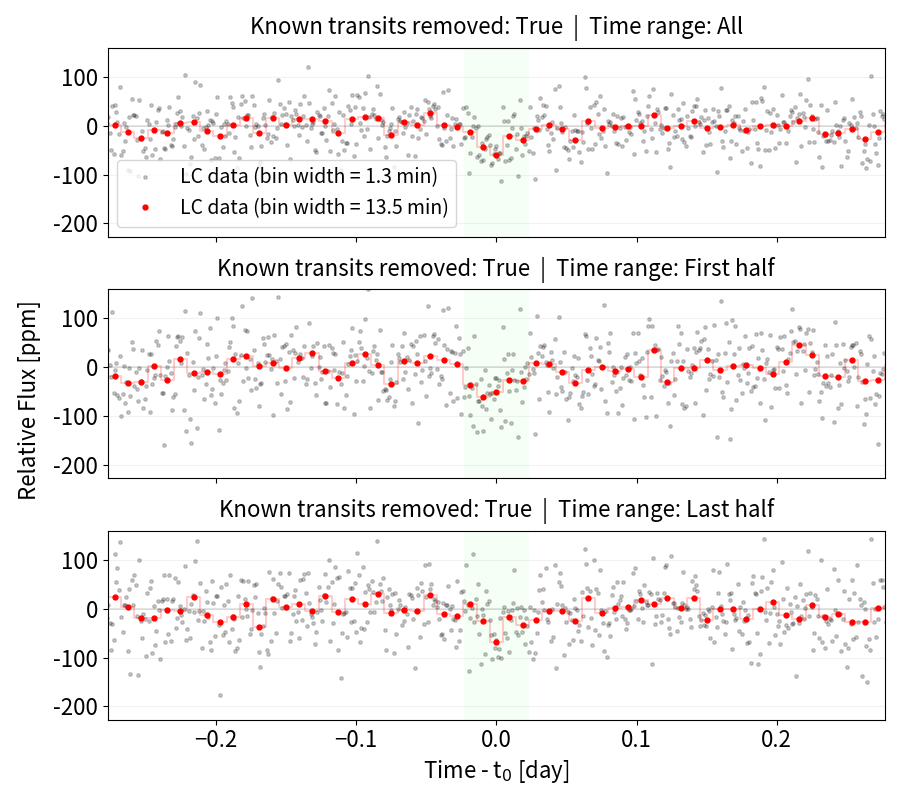}
    \caption{Left-half and Right-half Transit Check for candidate Kepler-963c. This test verifies that the transit signals of the new candidates remain consistent across both subsets of the light curve time sequence.}
    \label{fig:8832512_first_last_half}
\end{figure}

\textbf{5. Left-half and Right-half Transits Check}\newline
Through the left-right transit examination, we ensured consistency in the transit signals across the duration of the \textit{Kepler} mission for all of our candidates, mixing \textit{Kepler} seasons. Figure \ref{fig:8832512_first_last_half} provides a visualization of this left-right transit assessment for Kepler-963c.

\textbf{6. Column Anomaly Check}\newline
This test assesses potential interference linked to the "column anomaly" — a scenario where a star with varying luminosity, situated on the same CCD columns as the primary target, may introduce unwanted signals during the readout process \citep{Coughlin_2016}.
During its primary mission, the \textit{Kepler} spacecraft underwent a 90-degree roll approximately every 90 days to align its solar panels with the Sun, causing stars to be repositioned on different segments of the CCD array \citep{Coughlin_2016}. Consequently, contamination signals might be more pronounced in certain positions on the CCD array than others. Our assessment revealed no evidence of column anomalies in any of the candidate data.

\textbf{7. Bootstrap Validation}\newline
We conducted a bootstrap validation akin to the one used in \citet{Shahaf_2022}. For each candidate's raw light curve, we temporally shifted the light curve of each quarter by adding a uniformly random time offset in the range from 0 to the detected transit period specific to the candidate. Following this temporal perturbation, we processed the time-shifted light curve using our GPU-based folding and CNN processing pipeline to find the most prominent periodogram peak and calculated its SNR. This procedure was iterated ten times for each target. We identified the most significant remaining signal in the time-shifted light curve, computed its SNR, and then calculated the mean SNR and corresponding standard deviation across the ten iterations. In setting our validation criteria, we required that the SNR associated with the detected transit exceed four times the standard deviation derived from the ten bootstrap iterations. All of our candidates met this bootstrap validation criterion.

\textbf{8. Inversion Validation}\newline
To further distinguish the transit signals from non-planetary sinusoidal or similar periodic signals, we conducted an inversion test by reversing flux values around their nominal baseline, effectively converting the dips of a light curve into peaks, and vice versa \citep{Adams_2021}. Since our pre-processing step ensures that the average flux value of a detrended light curve is 1, the inversion process results in a new light curve with flux values equal to 2 minus their original values. The inverted light curve was then processed by our GPU folding and the CNN pipeline to detect the presence of a transit-like signal. In this test, no transit-like signals were detected in our candidates, therefore passing the inversion test.

\textbf{9. Outside Apertures Check}\newline
To ensure that the transit signals were not due to contamination from scattered background light, we conducted an outside apertures test \citep{Shallue_2018}. For each candidate, we generated light curves using pixels located outside the optimal aperture provided by \textit{Kepler}. These light curves were then folded based on their respective detected transit periods. Should the transit signals from the background be notably stronger than those from the on-target pixels, it would indicate potential contamination. Our test results showed no significant signals, confirming that the transit signals originated at or near the target star.

\begin{figure}
    \centering
    \includegraphics[scale = 0.2]{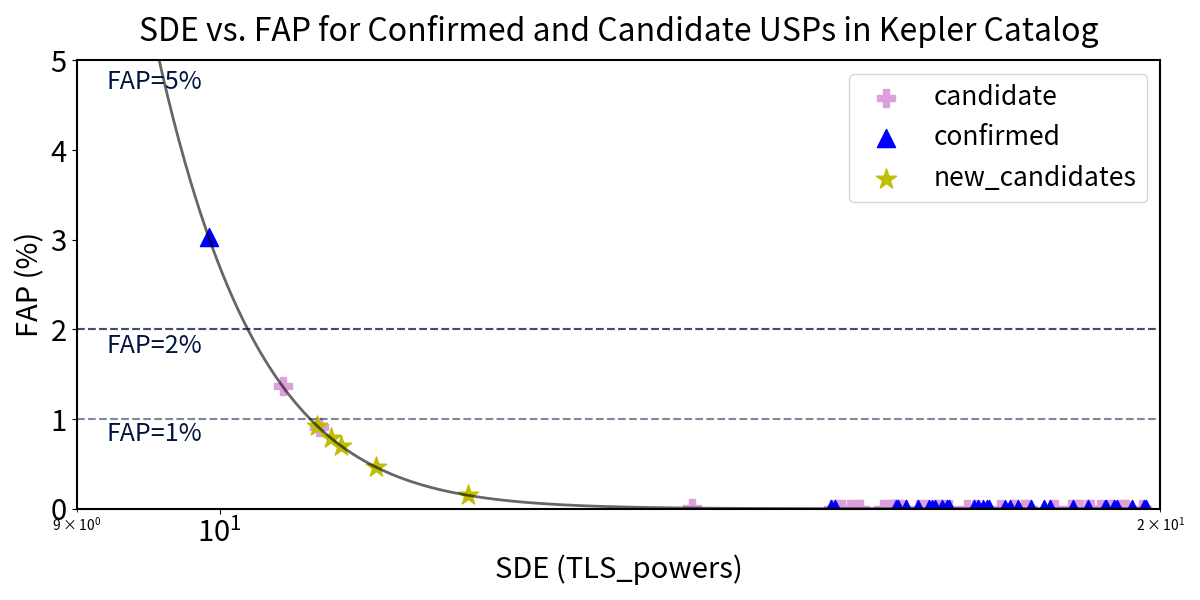}
    \caption{FAP check. The FAP vs. Signal Detection Efficiency (SDE) curve is plotted, with the FAPs for the confirmed and candidate USP exoplanets in the \textit{Kepler} catalog marked in triangle (blue) and cross (purple), respectively. The FAPs for our new candidates are marked with stars (yellow).}
    \label{fig:sde_fap_conf_cand_usps_kepler}
\end{figure}

\textbf{10. False Alarm Probability (FAP) Check}\newline
In our analysis, we conducted rigorous FAP testing to rule out data noise (random and systematic) influences on transit detection. We employed standard methods commonly used in transit analysis, such as those employed by the \textit{Kepler} team \citep{KeplerManual2016} and reinforced by recent research on \textit{Kepler} data \citep{Weiss2024Kepler}, and integrated into the Transit Least Squares (TLS) \citep{Hippke_TLS_2019} Python package. Our FAP assessment comprised two parts: one utilizing synthesized white noise light curves for baseline comparison, and another using the actual candidate light curves. The former established a benchmark, confirming that our candidates' FAP values align with those of \textit{Kepler}'s confirmed and candidate exoplanets where FAP is at or below 1\% except one confirmed one and one candidate. As illustrated in Figure \ref{fig:sde_fap_conf_cand_usps_kepler}, our candidates' FAPs fall well within this accepted range. The latter test, applied to each candidate's own light curve, demonstrated a dominance of white noise over red noise, suggesting that the detected transits are unlikely artifacts of random noise.

The TLS algorithm and its GPU-accelerated counterpart, GPU Transit Least Squares (GTLS) \citep{Hu2024GTLS}, were pivotal in our methodology, serving as the backbone for the evaluation of light curve data for the FAP analysis. \rt{Upon input of a light curve, GTLS generates a spectrum of Signal Detection Efficiency (SDE) scores, defined by the BLS method outlined in \citet{Kov_cs_2002}}, each corresponding to a unique orbital period. The utility of SDEs in FAP calculations is twofold: they are both noise-normalized and deterministic. This characteristic is crucial, given the diverse SNRs encountered in light curves. Noise normalization ensures that the SDE metric is consistent across varying SNR levels, facilitating the development of a universal model for the FAP relationship. Additionally, the deterministic nature of SDEs, derived directly from empirical light curve data, allows for consistent SDE values for a specific transit signal across different light curves, assuming comparable SNRs. This consistency is achieved through precise configuration of the GTLS settings, reinforcing the reliability of SDE as a robust metric in our analysis.

\begin{figure}
    \centering
    \includegraphics[scale = 0.48]{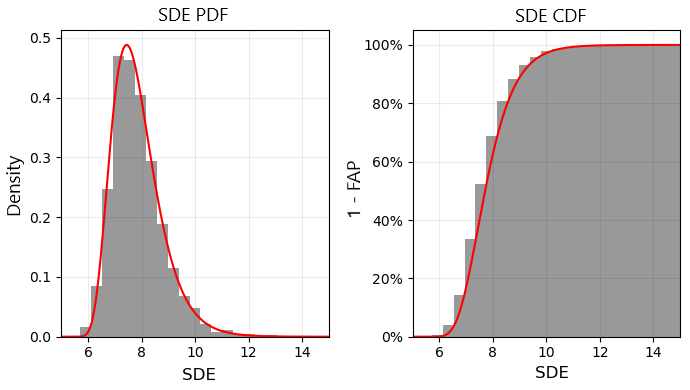}
    \caption{Production of the SDE-FAP relationship model using 10,000 synthetic light curves created from white noise. Using GTLS, we obtain the maximum SDE of each light curve and then create the histogram on the left to show the SDE Probability Density Function (PDF). This SDE PDF is fitted with a log-normal distribution (solid red line). The Cumulative Distribution Function (CDF) of this distribution is shown on the right, providing the 1-FAP at each SDE.}
    \label{fig:wn_sde_pdf_cdf}
\end{figure}

The construction of the FAP statistic began with the generation of 10,000 synthetic light curves, each consisting of 50,000 flux data points. These points were paired with a corresponding time array to represent observation times, with each time step mirroring \textit{Kepler}'s 29.4-minute observation interval. \rt{Measurements reveal that the SNR of \textit{Kepler} light curves typically ranges between 1000 and 4000. In accordance with this, the data points were drawn from a Gaussian distribution, with adjustments made prior to sampling to ensure the light curves maintained a mean flux of 1 and a randomly assigned SNR within the 1000 to 4000 range.} These synthetic light curves were then analyzed using the GTLS algorithm, which yielded SDE scores across specified orbital periods. Our search, focused on USPs, required the GTLS processing to be confined to an orbital period range of  [0.2 1] days. The maximal SDE value extracted from each light curve formed a dataset of 10,000 SDEs. From this, we constructed a Cumulative Distribution Function (CDF), delineating the relationship between SDE and 1-FAP. Further analysis involved fitting the Probability Distribution Function (PDF) of these SDEs with a log-normal distribution, resulting in an exemplary fit (R-squared = 0.997, $\chi^2_{\mathrm{reduced}}$ = 2.4, P-value = 0.038). This log-normal model of the PDF was then transformed into a CDF model, establishing a smooth and continuous correlation between SDE and 1-FAP, as depicted in Figure \ref{fig:wn_sde_pdf_cdf}. Finally, we calculated the FAP values by subtracting the 1-FAP percentages from unity, thereby deriving the definitive SDE-FAP relationship model for our study.

With the SDE-FAP model solved, we could then obtain FAP values for our candidates by running the detrended light curves through GTLS to obtain the SDE of the candidate transit signal, and then evaluating the SDE-FAP model to obtain the FAP estimate.

\begin{figure}
    \centering
    \includegraphics[scale = 0.44]{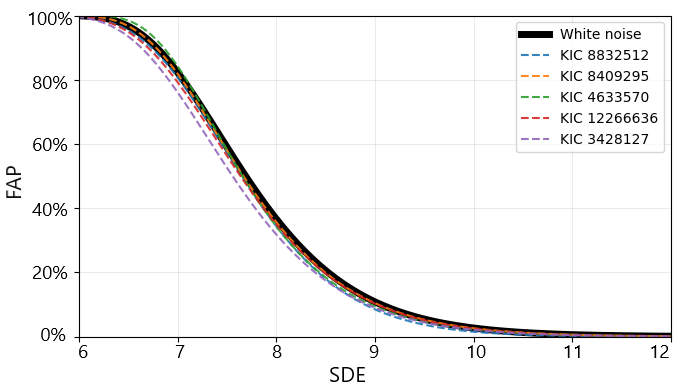}
    \caption{Comparison between the SDE-FAP model produced from the synthetic white noise light curves and the models generated by bootstrapping the light curves of our USP candidates. Notably, there is evident variation between the models when the FAP exceeds 5\%. However, this variation is largely reduced in the 0\% to 1\% FAP range, which is our primary focus for selection criteria. Remarkably, the white noise model sets the most stringent threshold, demanding an SDE of 10.69 to achieve a 1\% FAP. The other models, presented in the sequence corresponding to the plot legend, require SDEs of 10.24, 10.68, 10.60, 10.62, and 10.51 to reach the same 1\% FAP level.}
    \label{fig:combo_sde_fap}
\end{figure}

In the final verification step of the SDE-FAP relationship model, we used the light curves of our candidates to produce a new SDE-FAP model for each candidate. This step was taken to ensure that the model still holds true for these targets which would indicate that red noise was not dominant. To do this, for each light curve, we removed all known transits then detrended. Then each light curve was bootstrapped to produce 2,000 new light curves. Through the bootstrapping technique, we permuted the flux values, effectively randomizing their temporal sequence. After using GTLS to obtain SDEs and then fitting the PDF of the SDE maximums with a log-normal distribution model, we derived a SDE-FAP relationship model for each candidate. These models are compared to the white noise SDE-FAP model in Figure \ref{fig:combo_sde_fap}.

\begin{figure}
    \centering
    \includegraphics[scale = 0.47]{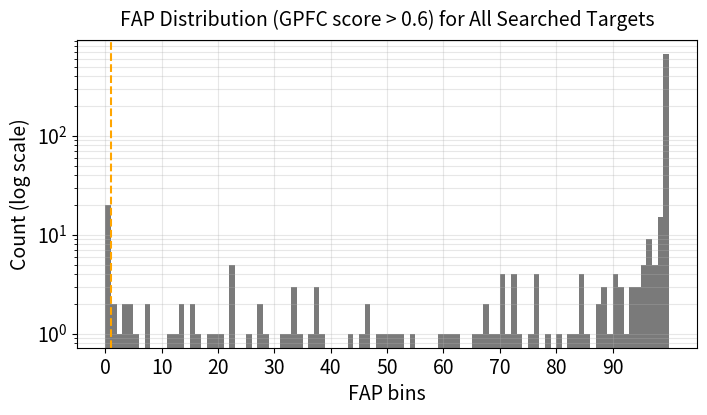}
    \caption{Distribution of FAPs for all of searched targets with GPFC scores exceeding 0.6. Our search covered hosts containing confirmed or candidate orbiting USPs in the KOI catalog. A dashed line marks the cutoff threshold at an FAP of 1\%. }
    \label{fig:FAP_distr_GPFC_06}
\end{figure}

Figure \ref{fig:FAP_distr_GPFC_06} presents the FAP distribution for targets exhibiting GPFC scores greater than 0.6. \rt{In our examination of \textit{Kepler} data, we found that nearly all confirmed and candidate USPs have FAP values below 1\%, except for one instance in each category—confirmed and candidate USPs—where the FAP values are above 1\%.} Therefore we adopted FAP values under 1\% as the threshold for selecting candidates for additional validation. The FAP values for our identified candidates, detailed in Table~\ref{tab:robust_candidates}, all fall below this 1\% threshold.

After conducting the initial nine vetting tests mentioned above, we narrowed down the field from approximately 40 candidates to 12. Following the FAP check, this number was further refined, finalizing our list to 5 candidates.

\rt{\subsection{Candidate Validation Results}}

\rt{To validate our planet candidates as true detections, the probabilistic validation method introduced by \citet{Morton_2016ApJ} was applied, in which candidates with false positive probabilities (FPP) <1\% of being astrophysical false positives are considered validated planets.}

\begin{figure} 
    \centering
    \includegraphics[scale = 0.65]{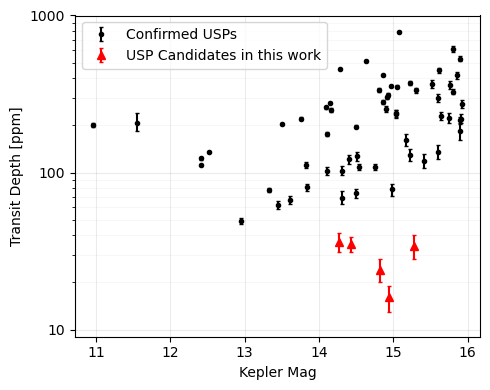}
    \caption{\rt{This graph displays transit depths against \textit{Kepler} magnitudes, differentiating between USP KOIs confirmed through Vespa and the USP candidates identified in this research. There is a marked separation between the datasets, reflecting the higher SNR of Vespa-confirmed USPs with FPPs under 1\%. Due to their lower SNRs, the USP candidates identified in this study did not meet the requisite threshold for Vespa analysis.}}
    \label{fig:usp_depth_vs_kpmag}
\end{figure}

\rt{One way to calculate astrophysical FPP is to use statistical tools like Vespa. However, \citet{Morton_2016ApJ} advises caution in interpreting Vespa results for low SNR candidates and due to their lower SNRs, the USP candidates identified in this study did not meet the requisite threshold of Multiple Event Statistic (MES) of greater than 10 for Vespa analysis (Table \ref{tab:false_positive_probabilities}). Figure \ref{fig:usp_depth_vs_kpmag} illustrates the poor SNRs of our USPs compared to USPs confirmed by Vespa.}

\begin{figure} 
    \centering
    \includegraphics[scale = 0.62]{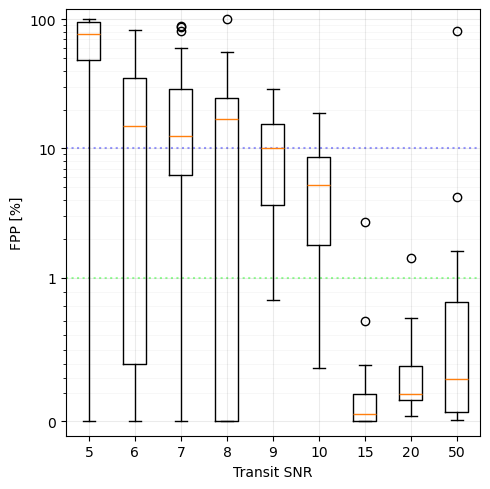}
    \caption{\rt{Boxplot of Vespa FPP for injected synthetic transits across a range of transit SNR. The X-axis displays transit SNR values on a non-linear scale. The whiskers of the boxplot extend to the 0th and 100th percentiles, with the box boundaries representing the 25th and 75th percentiles, and the orange horizontal bars indicating the median (50th percentile) of the data. \rtt{The circles mark outliers.} Synthetic transits, generated by the Batman package \citep{batman}, injected into real \textit{Kepler} light curves were used to investigate the impact of transit SNR on FPP outcomes. Notably, FPP variability is pronounced at transit SNRs $\leq$10. The five USP candidates presented in this work have both transit SNR and Multiple Event Statistic (MES) below 10, underscoring the unsuitability of Vespa's analysis for these particular candidates due to their low SNR.}}
    \label{fig:vespa_snr_vs_fpp}
\end{figure}

\rt{An empirical test was further conducted to gauge Vespa's FPP calculation reliability for low SNR transits. Synthetic transits with varied SNRs, created using the Batman package \citep{batman}, were injected into our USPs' light curves. Figure \ref{fig:vespa_snr_vs_fpp} presents the results, illustrating FPP variability as a function of transit SNR. Specifically, at SNRs below 10, the FPP demonstrated a wide range, extending from approximately 0\% to 100\% in various tests at each SNR. These findings corroborate the data quality and analysis reliability concerns previously raised by \citet{Morton_2016ApJ}.}

\rt{The other way to calculate astrophysical FPP is to use the statistical method based on planet multiplicity suggested by \citet{Lissauer_2012ApJ}. Each of our planet candidates has been identified within systems that are already documented as hosting at least one additional planet (such as KIC 4633570, KIC 8832512, KIC 12266636 and KIC 8409295) or candidate (such as KIC 3428127). \citet{Ragozzine_2010}, and \citet{Lathm_2011} have stated that KOIs related to stars harboring multiple planet candidates have a significantly lower probability of being astrophysical false positives compared to KOIs linked to stars with a single candidate. Given that the stars associated with our candidates are already known to host at least one exoplanet or planet candidate, the probability of such stars both hosting an exoplanet and having a very nearby background binary star is greatly reduced over that for a random field star. This strengthens the credibility of our discoveries as true planetary candidates, rather than mis-identifications of background binary stars.}

\begin{table}
    \centering
    \caption{FPP metrics for the five candidates. Multiple-event statistic (MES), used in \citet{Morton_2016ApJ}, is a metric equivalent to SNR. Signal Detection Efficiency (SDE) is a score generated by the TLS approach \citep{Hippke_TLS_2019} which analyzes transit features particularly suited for small planets, and we applied the accelerated GTLS \citep{Hu2024GTLS} as an efficient way to calculate these scores. FPP differs in that we estimate the probability of an astrophysical false positive based on known companions to the candidates orbiting the host star. }
    
    \begin{tabular}{ l | l | l | l }
    \hline
    \hline
    \textbf{Target} & \textbf{MES} & \textbf{SDE} & \textbf{FPP} \\
    \hline
    KIC 4633570 & 6.6 & 10.94 & 0.014\% $\pm$ 0.02\% \\
    KIC 8832512 & 6.4 & 12.17 & 0.12 $\pm$ 0.01 \% \\
    KIC 12266636 & 5.6 & 10.85 & 0.12 $\pm$ 0.01 \% \\
    KIC 8409295 & 4.1 & 11.22 & 0.12 $\pm$ 0.01 \% \\
    KIC 3428127 & 4.3 & 10.75 & 0.41\% $\pm$ 0.02\% \\
    \hline
    \end{tabular}
    \label{tab:false_positive_probabilities}
\end{table}

\rt{Our selection process targeted hosts from the KOI catalog identified with planets of either \rtt{confirmed or candidate disposition. Specifically, the host stars of four of our detections KIC 4633570, KIC 8832512, KIC 12266636, and KIC 8409295 each have at least one other confirmed planet known to orbit the star. KIC 4633570 currently hosts two confirmed planets, KIC 12266636 hosts one confirmed and one candidate planet, and KIC 8832512 and KIC 8409295 both host a single confirmed planet.} These planets were validated using Morton's statistical method \citep{Morton_2016ApJ}, which validates candidates with Vespa FPP below 1\%. Their star and planet radii were revised using Gaia DR2 data \citep{Berger_2018ApJ}. Moreover, KIC 4633570 and KIC 12266636 were subjected to difference image analysis to detect potential background eclipsing binaries \citep{Batalha_2013ApJ}. Notably, KIC 4633570 also underwent further validation for its multi-planet candidates, including FPP checks below 1\% and centroid motion analysis \citep{Rowe_2014ApJ}. The host star KIC 3428127, with its planet candidate disposition, has yet to undergo validation. Follow-up observations included adaptive optics (AO) imaging for all five host stars, with spectroscopic observations conducted on all except KIC 8832512. Previous work showed that the overwhelming majority of \textit{Kepler} candidates in multiple transiting systems are true planets \citep{Lissauer_2012ApJ} and has used this principle to validate \textit{Kepler} planets in multi-planet systems \citep{Rowe_2014ApJ}.}

\citet{Lissauer_2012ApJ} provides formulas to quantify the increased probability of planethood of candidates in a multi-planet system, which we applied to estimate the FPP of our detections. Given a base probability $P$ of planethood for a single candidate, Equation 4 in \citet{Lissauer_2012ApJ} gives the probability of 1 planet and 1 FP, Equation 6 gives the probability of 2 planets and 1 FP, and Equation 8 gives the FPP accounting for multiplicity for a detection in a two candidate system. \rtt{\citet{Fressin_2013ApJ} conducted simulations and a statistical analysis to estimate the false positive rate (FPR) across the Kepler catalog. This FPR value is broken down into varying FPRs among planets of different size groups (grouped by planetary radii). We estimated an updated cumulative FPR by scaling the FPRs of each size group found by \citet{Fressin_2013ApJ} to current KOI candidate catalog totals. With 546 Earths ($<1.25$ $R_{\oplus}$, FPR 12.3\% $\pm$ 3.0 \%), 536 Super Earths (1.25 - 2 $R_{\oplus}$, FPR 8.8\% $\pm$ 1.9\%), 430 Small Neptunes (2 - 4 $R_{\oplus}$, FPR 6.7\% $\pm$ 1.1\%), 74 Large Neptunes (4 - 6 $R_{\oplus}$, FPR 15.9\% $\pm$ 3.5 \%), and 292 Giants ($>$6 $R_{\oplus}$, FPR 17.7\% $\pm$ 2.9 \%) among a total of 1878 candidates in the current catalog, this gives a cumulative FPR of 11 $\pm$ 1.0\%.}

\rt{Our search was conducted among the pool of KOIs in the confirmed and candidate catalogs, with  a total of 4716 targets. Following \citet{Fressin_2013ApJ}, we set \rtt{$P$, the probability of planethood for a single planet in our search sample}, to $n_p / n_c = 95.5\% \pm 0.5\%$, where $n_p$ includes the confirmed and true candidates under the estimated FPR of 11 $\pm$ 1\% and $n_c$ is the total confirmed and candidate KOI. Thus, for Kepler-963c, Kepler-879c, and Kepler-1498c, the probability of being a FP in a system with 1 confirmed detection is estimated to be 0.12\% $\pm$ 0.01\%. For Kepler-158d with two confirmed companions, the FPP is estimated to be 0.014\% $\pm$ 0.02\%. For KOI-4978.02, with only a candidate companion planet, we use the conservative value of $P = 89\% \pm 1.0\%$ directly to estimate an FPP of 0.41\% $\pm$ 0.02\%.}

\rt{These results indicate that the probability of astrophysical false positives for all five candidates is below 1\%. Furthermore, our FAP test using the TLS method shows that the likelihood of \rtt{each of these five candidate signals} being produced by random or systematic data noise is also below 1\%. Therefore, our analysis demonstrates that the five identified Ultra-Short Period (USP) candidates are statistically validated.}

\subsection{Transit Fitting}

\rtt{The five USPs were fit with the \citet{Mandel_2002} model, providing estimates of the planet and orbital parameters of the candidates. The parameter space was explored using the Markov Chain Monte Carlo (MCMC) algorithm \citep{Goodman2010EnsembleSW} through the PyTransit \citep{Parviainen_Pytransit_2015} package.}

\rtt{The basis set of parameters was orbital period ($p$), transit center ($t_c$), impact parameter ($b$), radius ratio ($k$), eccentricity ($e$), argument of periapsis ($\omega$), stellar density ($\rho$), and the limb darkening coefficients $(q_1, q_2)$. To determine which of these parameters could be solved with reasonable confidence, testing with synthetic transits injected into real light curves was performed. In this test, a random KOI was chosen and its PDCSAP light curve loaded. A random set of orbital parameters was generated then used to generate a synthetic transit model. This model was then multiplied into the Kepler light curve. The light curve was then subjected to the same detrending, transit masking, and MCMC analysis that would be applied to the USP candidates. With this process, 50 samples were generated and measured.}

\begin{figure}
    \centering
    \includegraphics[scale = 0.65]{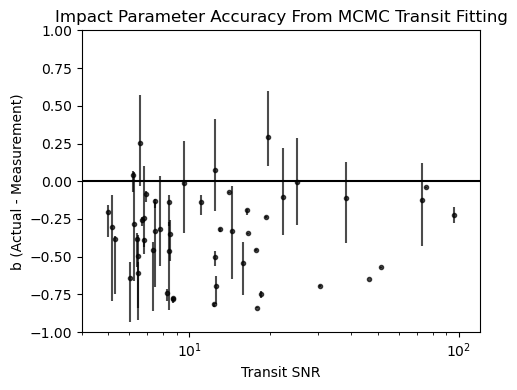}
    \caption{Difference of actual and measured impact parameter of 50 light curves injected with synthetic transits. All transits were from simulated USP (0.1 to 1.0 day orbital periods) planets. Given the inaccuracy of impact parameter estimation by MCMC analysis, we opted to not solve for it and set the impact parameter to zero.}
    \label{fig:mcmc_test_delta_b_snr}
\end{figure}

\rtt{Results (Fig. \ref{fig:mcmc_test_delta_b_snr}) show that planets with orbital periods of less than 1 day and a cumulative transit SNR of less than about 50 exhibited extreme difficulty in measuring their impact parameter, eccentricity, argument of periapsis, and limb darkening coefficients. Due to this, when fitting our candidates, we set the impact parameter, eccentricity, and argument of periapsis to zero while fixing the limb darkening parameters to those estimated by \citet{Claret_Bloemen_2011}. With these settings, these parameters are not solved during fitting. By setting the impact parameter ($b$) to zero, the orbit inclination angle is effectively set to 90 degrees. The stellar density was also set to that reported in the KOI catalog. The following is a breakdown of the basis set of parameters and the configuration used as priors in the MCMC transit fitting.}

\begin{itemize}
    \setlength{\itemsep}{1mm} 
    \setlength{\parskip}{0mm} 
    \setlength{\itemindent}{0em}
    \item \textbf{Orbital period} ($p$) \\
    Mean: determined in the transit search and verified with GTLS and a preliminary least-squares transit fitting \\
    Uncertainty: $4 \times 10^{-6}$ day; a typical uncertainty for USPs in Kepler data.

    \item \textbf{Transit center} ($t_c$) \\
    Mean: determined by GTLS and a preliminary least-squares transit fitting \\
    Uncertainty: 0.004 day; a typical uncertainty for USPs in Kepler data.

    \item \textbf{Impact parameter} ($b$) -- Set to 0. Effectively sets the inclination angle to 90 degrees.

    \item \textbf{Radius ratio} ($k$) \\
    Mean: estimated by a preliminary least-squares transit fitting \\
    Uncertainty: 0.005; Liberal estimate of uncertainty to ensure a wide range of radius ratios are evaluated.

    \item \textbf{Eccentricity} ($e$) -- Set to 0.

    \item \textbf{Argument of periapsis} ($\omega$) -- Set to 0.

    \item \textbf{Stellar density} ($\rho$) -- Set to the stellar density calculated using the estimated stellar mass and radius reported in Kepler’s KOI catalog.

    \item \textbf{Limb darkening coefficients} $(q_1, q_2)$ -- Set to the estimated limb darkening coefficients reported in the Kepler catalog. Note that Kepler reports the $(u_1, u_2)$ version which is then converted to the $(q_1, q_2)$ representation, as required by PyTransit.
\end{itemize}

Additionally, the radial velocity amplitude of the planet is estimated using the planetary parameters and Equation \ref{planet_RV}, where $M_p$ represents the planet mass in Jupiter masses, $M_\odot$ is the stellar mass in solar masses, $i$ is the inclination angle in radians, and $P$ is the orbital period in years. Stellar mass (in M$_\odot$) and radius (in R$_\odot$) used in the calculations are sourced from the NASA Exoplanet Archive, with data derived from the Dartmouth Stellar Evolution Database (DSEP).   Given that all the five USPs are assumed to be rocky, we use Earth's density as the basis for estimating \rtt{the planet's mass in the expression for radial velocity} \citep{Luque_Palle_2022Sci}.

\rt{
\begin{equation}
    \label{planet_RV}
    RV = \frac{28.4 M_p sin(i)}{P^{1/3} M_\odot^{2/3}}
\end{equation}
}

\begin{figure}
  \centering
  \includegraphics[scale = 0.65]{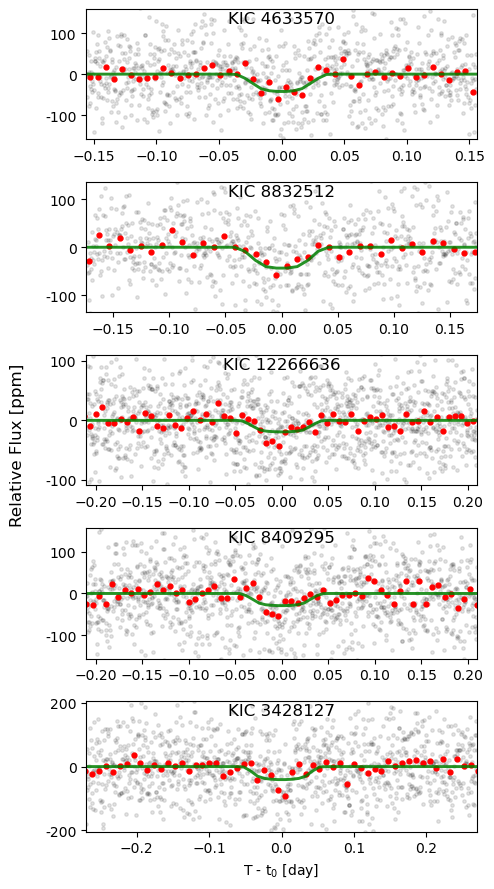}
  \caption{Markov Chain Monte Carlo (MCMC) fitting plot displaying the highest likelihood estimates for the five USPs. The flux data are folded with the transit period and binned into 1.9-minute bins (grey dots), and 3.1-4.5 minute intervals (red dots). The black curve represents the best-fitting transit model, with the grey shaded area in the transit window indicating the model uncertainty.}
  \label{fig:all_mcmc_fittings}
\end{figure}

\begin{figure} 
    \centering
    \includegraphics[scale = 0.37]{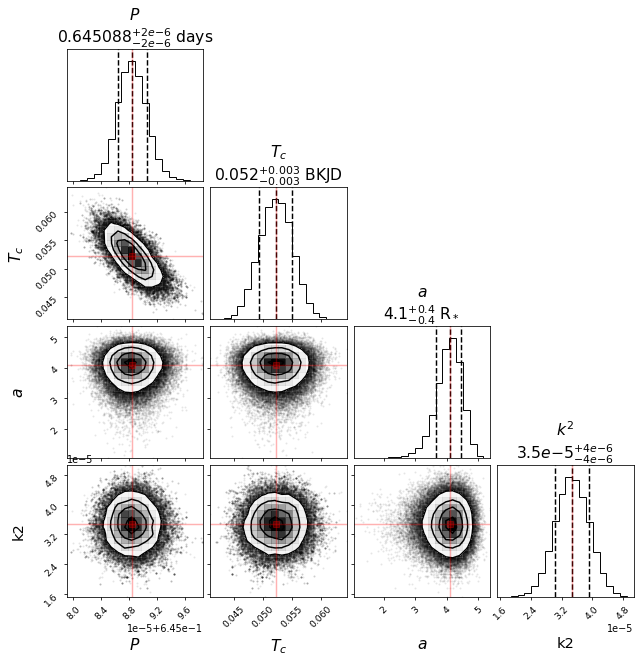}
    \caption{MCMC fitting corner plot for USP Kepler-158d. It shows the parameters including central transit time ($T_c$), orbital period ($P$), radius ratio squared ($k^2$), scaled semi-major axis ($a$). The red lines represent the most likely values, while the dashed lines reflect the one-sigma uncertainty estimations.}
    \label{fig:4633570_mcmc_corner_plot}
\end{figure}

 \begin{table*}
     \caption{Stellar and derived planetary parameters of the five new USPs. The stellar effective temperature $\mathrm{T}_{\mathrm{eff}}$ (in Kelvin) is retrieved from the Gaia Archive DR2. \rt{Stellar mass (in M$_\odot$), radius (in R$_\odot$), and density (in g/cm$^3$) are obtained from the NASA Exoplanet Archive, sourced from the Dartmouth Stellar Evolution Database (DSEP).} Planetary parameters, such as the time of transit ($T_c$, in BKJD), transit duration (in days), orbital period (in days), semi-major axis (in stellar radii), radius (in Earth raedii), \rt{and the estimated radial velocity amplitude (in m/s)}, are derived from MCMC fitting.}
    
    \renewcommand{\arraystretch}{1.4} 
    \begin{tabular} { p{4cm} p{2.1cm} p{2.1cm} p{2.1cm} p{2.1cm} p{2.1cm}}
     \hline
     \hline
     \textbf{Stellar Parameters} & KIC 4633570 &KIC 8832512 &KIC 12266636 &KIC 8409295 &KIC 3428127\\
     \hline
    $\mathrm{T}_{\mathrm{eff}}$ [K] & ${4895.97}$ [K] & ${5169.47}$ [G] & ${5456.75}$ [G] & ${5565.53}$ [G] & ${5050.28}$ [G]\\
    Mass  [M$_\odot$]& 
    ${0.656}_{{\:\scriptscriptstyle-{0.049}}}^{{\:\scriptscriptstyle+{0.073}}}$ & 
    ${0.821}_{{\:\scriptscriptstyle-{0.065}}}e^{{\:\scriptscriptstyle+{0.12}}}$ &
    ${0.974}_{{\:\scriptscriptstyle-{0.104}}}^{{\:\scriptscriptstyle+{0.127}}}$ & 
    ${0.977}_{{\:\scriptscriptstyle-{0.062}}}^{{\:\scriptscriptstyle+{0.05}}}$ & 
    ${0.859}_{{\:\scriptscriptstyle-{0.065}}}^{{\:\scriptscriptstyle+{0.126}}}$\\
    
    Radius  [R$_\odot$]& 
    ${0.668}_{{\:\scriptscriptstyle-{0.059}}}^{{\:\scriptscriptstyle+{0.054}}}$ & 
    ${0.928}_{{\:\scriptscriptstyle-{0.147}}}^{{\:\scriptscriptstyle+{0.239}}}$ & 
    ${1.022}_{{\:\scriptscriptstyle-{0.155}}}^{{\:\scriptscriptstyle+{0.288}}}$ & 
    ${0.957}_{{\:\scriptscriptstyle-{0.066}}}^{{\:\scriptscriptstyle+{0.132}}}$ & 
    ${0.966}_{{\:\scriptscriptstyle-{0.133}}}^{{\:\scriptscriptstyle+{0.385}}}$\\

    Stellar Density [g/cm$^3$] & 
    ${3.1}_{{\:\scriptscriptstyle-{0.9}}}^{{\:\scriptscriptstyle+{0.9}}}$ & 
    ${3.2}_{{\:\scriptscriptstyle-{0.8}}}^{{\:\scriptscriptstyle+{0.8}}}$ & 
    ${1.6}_{{\:\scriptscriptstyle-{0.9}}}^{{\:\scriptscriptstyle+{0.8}}}$ & 
    ${1.6}_{{\:\scriptscriptstyle-{0.6}}}^{{\:\scriptscriptstyle+{0.9}}}$ & 
    ${1.1}_{{\:\scriptscriptstyle-{0.5}}}^{{\:\scriptscriptstyle+{0.8}}}$\\


    \hline
    \hline
    \textbf{Derived Planet Parameters}&Kepler-158d & Kepler-963c & Kepler-879c & Kepler-1489c & KOI-4978.02\\
    \hline
    T$_c$ [BKJD] & 
    ${0.052}_{{\:\scriptscriptstyle-{0.003}}}^{{\:\scriptscriptstyle+{0.003}}}$ & 
    ${-0.099}_{{\:\scriptscriptstyle-{0.003}}}^{{\:\scriptscriptstyle+{0.003}}}$ & 
    ${0.206}_{{\:\scriptscriptstyle-{0.003}}}^{{\:\scriptscriptstyle+{0.003}}}$ & 
    ${-0.040}_{{\:\scriptscriptstyle-{0.003}}}^{{\:\scriptscriptstyle+{0.003}}}$ & 
    ${0.099}_{{\:\scriptscriptstyle-{0.003}}}^{{\:\scriptscriptstyle+{0.003}}}$\\

  Transit duration [days] & 
    ${0.052}_{{\:\scriptscriptstyle-{0.007}}}^{{\:\scriptscriptstyle+{0.007}}}$ & 
    ${0.058}_{{\:\scriptscriptstyle-{0.005}}}^{{\:\scriptscriptstyle+{0.005}}}$ & 
    ${0.07}_{{\:\scriptscriptstyle-{0.03}}}^{{\:\scriptscriptstyle+{0.03}}}$ & 
    ${0.07}_{{\:\scriptscriptstyle-{0.02}}}^{{\:\scriptscriptstyle+{0.02}}}$ & 
    ${0.09}_{{\:\scriptscriptstyle-{0.02}}}^{{\:\scriptscriptstyle+{0.02}}}$\\
     
    Orbital period [days] & 
    ${0.645088}_{{\:\scriptscriptstyle-{2e-06}}}^{{\:\scriptscriptstyle+{2e-06}}}$ &
    ${0.919783}_{{\:\scriptscriptstyle-{3e-06}}}^{{\:\scriptscriptstyle+{3e-06}}}$ &
    ${0.646716}_{{\:\scriptscriptstyle-{3e-06}}}^{{\:\scriptscriptstyle+{3e-06}}}$ &
    ${0.680741}_{{\:\scriptscriptstyle-{4e-06}}}^{{\:\scriptscriptstyle+{4e-06}}}$ &
    ${0.941967}_{{\:\scriptscriptstyle-{3e-06}}}^{{\:\scriptscriptstyle+{3e-06}}}$\\
    
    Semi-major axis [R$_\star$] & 
    ${4.0}_{{\:\scriptscriptstyle-{0.4}}}^{{\:\scriptscriptstyle+{0.4}}}$ & 
    ${5.2}_{{\:\scriptscriptstyle-{0.4}}}^{{\:\scriptscriptstyle+{0.4}}}$ & 
    ${3.1}_{{\:\scriptscriptstyle-{0.7}}}^{{\:\scriptscriptstyle+{0.7}}}$ & 
    ${3.3}_{{\:\scriptscriptstyle-{0.5}}}^{{\:\scriptscriptstyle+{0.5}}}$ & 
    ${3.6}_{{\:\scriptscriptstyle-{0.6}}}^{{\:\scriptscriptstyle+{0.6}}}$\\



    Radius [R$_\oplus$]&
    ${0.43}_{{\:\scriptscriptstyle-{0.05}}}^{{\:\scriptscriptstyle+{0.05}}}$ & 
    ${0.6}_{{\:\scriptscriptstyle-{0.2}}}^{{\:\scriptscriptstyle+{0.2}}}$ & 
    ${0.4}_{{\:\scriptscriptstyle-{0.1}}}^{{\:\scriptscriptstyle+{0.1}}}$ & 
    ${0.51}_{{\:\scriptscriptstyle-{0.08}}}^{{\:\scriptscriptstyle+{0.08}}}$ & 
    ${0.7}_{{\:\scriptscriptstyle-{0.1}}}^{{\:\scriptscriptstyle+{0.1}}}$\\
    
    Estimated RV amplitude [m/s]&
    ${0.10}_{{\:\scriptscriptstyle-{0.05}}}^{{\:\scriptscriptstyle+{0.05}}}$ & 
    ${0.16}_{{\:\scriptscriptstyle-{0.16}}}^{{\:\scriptscriptstyle+{0.16}}}$ & 
    ${0.09}_{{\:\scriptscriptstyle-{0.11}}}^{{\:\scriptscriptstyle+{0.11}}}$ & 
    ${0.52}_{{\:\scriptscriptstyle-{0.35}}}^{{\:\scriptscriptstyle+{0.34}}}$ & 
    ${1.2}_{{\:\scriptscriptstyle-{0.9}}}^{{\:\scriptscriptstyle+{0.9}}}$\\

    \hline
    \end{tabular}
    \label{tab:mcmc_parameters}
\end{table*}

 Figure \ref{fig:all_mcmc_fittings} shows the MCMC fitting plots for the five USPs. As an illustration, the corner plot for Kepler-158d is depicted in Figure \ref{fig:4633570_mcmc_corner_plot}. We present the parameters of the five USPs derived through MCMC fitting in Table \ref{tab:mcmc_parameters}.


As derived from MCMC fitting, Kepler-158d has a radius of approximately 0.43 R$_\oplus$. Hosted by KIC 4633570, a K-type main sequence star with a mass of 0.656 M$_\odot$ and a radius of 0.668 R$_\odot$, this new planet  turns KIC 4633570 into a three-planet multi-planetary system. Currently, the system hosts two confirmed planets: Kepler-158b (radius: 2.27 R$_\oplus$, orbital period: 16.71 days) and Kepler-158c (radius: 1.97 R$_\oplus$, orbital period: 28.55158 days)\footnote{https://exoplanetarchive.ipac.caltech.edu/}. Kepler-158d, with its short 0.645088-day orbit, is the system's smallest and innermost planet.

Kepler-963c possesses a radius of 0.6 R$_\oplus$. Its host, KIC 8832512, is a G-type main-sequence star with a mass of 0.821 M$_\odot$ and a radius of 0.928 R$_\odot$. The planet's addition makes KIC 8832512 a two-planet system, accompanying a previously confirmed planet, Kepler-963b with a radius of 2.85 R$_\oplus$ and a 9.97682-day orbit$^1$.

Kepler-879c  possesses a radius of 0.4 R$_\oplus$. It orbits KIC 12266636, a Sun-like G-type star (mass: 0.974 M$_\odot$, radius: 1.022 R$_\odot$). This addition turns KIC 12266636 into a three-planet system, alongside two other planets, Kepler-879b with a radius of 2.48 R$_\oplus$ and a 12.654896-day orbit and KOI-1522.02 with a radius of 1.02 R$_\oplus$ and a 33.3856176-days orbit$^1$.

Kepler-1489c possesses a radius of 0.51 R$_\oplus$. It orbits KIC 8409295, a Sun-like G-type main-sequence star (mass: 0.977 M$_\odot$, radius: 0.957 R$_\odot$). This addition supplements an already confirmed planet in the system, Kepler-1489b with a radius of 1.82 R$_\oplus$ and an 82.292865-day orbit$^1$, marking KIC 8409295 as a two-planet system.

Lastly, KOI-4978.02 possesses a radius of 0.7 R$_\oplus$. Its host, KIC 3428127, is a G-type main-sequence star characterized by a mass of 0.859 M$_\odot$ and a radius of 0.966 R$_\odot$. This addition complements another candidate planet. KOI-4978.01 with a radius of 2.75 R$_\oplus$ and an orbital period of 339.18998 days$^1$, making KIC 3428127 a two-planet system.

\section{Discussions}

In this section, we conduct a thorough analysis of our identified USPs, examining their distribution in the landscape of all USP planets discovered by the \textit{Kepler} survey. Additionally, we explore the implications of our findings concerning host star spectral types, the broader context of multi-planetary systems, and their alignment with prevailing theories of star formation.

\begin{table*}
    \caption{\rt{Comparisons of Kepler-band magnitudes, along with the spectral types of stars hosting the five smallest USPs in the NASA Confirmed Exoplanets Archive, are listed alongside details of our new USPs for reference. Notably three of our newly discovered  USPs rank among the five smallest confirmed USPs.}}  
    
    \begin{tabular}{ | l | c | c | c | c | c | }
        \hline
        \noalign{\vskip 4pt}
        \multicolumn{1}{c|}{\textbf{Host Star}} & \textbf{KIC} & \textbf{Stellar Type} & \textbf{Period} & \textbf{Radius} & \textbf{Kepler Magnitude} \\
        & & & \textbf{[days]} & \textbf{[R$_\oplus$]} & \\
        \hline
        \hline
        \multicolumn{6}{|c|}{\textbf{The Top Five Smallest NASA Confirmed USP Exoplanets}} \\
        \hline
        Kepler-1971b & 6592335 & M & 0.74 & 
        ${0.42}_{{\:\scriptscriptstyle-{0.08}}}^{{\:\scriptscriptstyle+{0.07}}}$ & 15.896 \\        
        Kepler-1579b & 7826620 & K & 0.85 & 
        ${0.59}_{{\:\scriptscriptstyle-{0.06}}}^{{\:\scriptscriptstyle+{0.04}}}$ & 15.220 \\
        Kepler-1351b & 6697756 & K & 0.92 & 
        ${0.60}_{{\:\scriptscriptstyle-{0.06}}}^{{\:\scriptscriptstyle+{0.05}}}$ & 14.107 \\
        Kepler-1566b & 5340878 & K & 0.54 & 
        ${0.68}_{{\:\scriptscriptstyle-{0.04}}}^{{\:\scriptscriptstyle+{0.11}}}$ & 14.311 \\
        Kepler-1067b & 8804845 & G & 0.76 & 
        ${0.7}_{{\:\scriptscriptstyle-{0.05}}}^{{\:\scriptscriptstyle+{0.23}}}$ & 14.305 \\        
        \hline
        \hline
        \multicolumn{6}{|c|}{\textbf{New Candidates}} \\
        \hline
        Kepler-879c & 12266636 & G & 0.646716 & 
        ${0.4}_{{\:\scriptscriptstyle-{0.1}}}^{{\:\scriptscriptstyle+{0.1}}}$ & 14.264 \\
        Kepler-158d & 4633570 & K & 0.645088 & 
        ${0.43}_{{\:\scriptscriptstyle-{0.05}}}^{{\:\scriptscriptstyle+{0.05}}}$ & 14.427 \\
        Kepler-963c & 8832512 & G & 0.919783 & 
        ${0.51}_{{\:\scriptscriptstyle-{0.08}}}^{{\:\scriptscriptstyle+{0.08}}}$ & 14.828 \\       
        KOI-4978.02 & 3428127 & G & 0.941967 & 
        ${0.6}_{{\:\scriptscriptstyle-{0.2}}}^{{\:\scriptscriptstyle+{0.2}}}$ & 14.943 \\
        Kepler-1489c & 8409295 & G & 0.680741 & 
        ${0.7}_{{\:\scriptscriptstyle-{0.1}}}^{{\:\scriptscriptstyle+{0.1}}}$ & 15.278 \\
        \hline
    \end{tabular}
    \label{tab:smallest_5_VK_mag}
\end{table*}

\subsection{Period Radius Distribution}

\begin{figure}
  \centering
  \begin{minipage}[c]{0.45\textwidth}
    \centering
    \subfloat{\includegraphics[width=\textwidth]
    {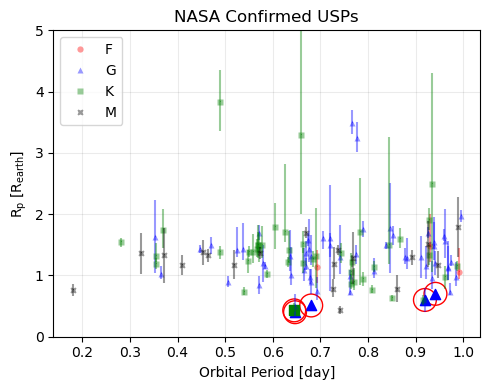}}\label{fig:pr_all_usp_nasa}
  \end{minipage}
  \vfill
  \vspace*{1em}
  \begin{minipage}[c]{0.45\textwidth}
    \centering
    \subfloat{\includegraphics[width=\textwidth]
    {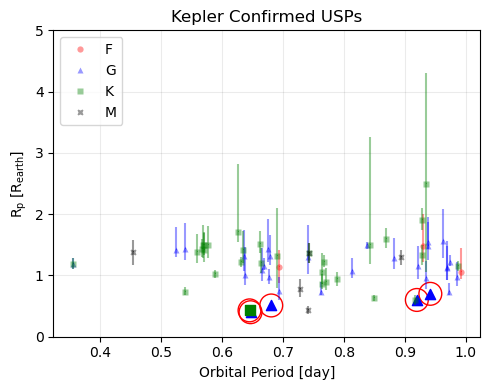}}\label{fig:pr_all_usp}
  \end{minipage}
  \caption{\rt{Period-Radius distribution of confirmed USPs in the NASA Exoplanet Archive. The upper plot illustrates the Period-Radius distribution of all USPs from the NASA Confirmed Exoplanet Archive, while the lower plot displays the same distribution for all confirmed USPs in the \textit{Kepler} KOI Catalog.} USPs orbiting host stars of types F, G, K, and M are color coded red, blue, green, and black, respectively. Our five new USPs are highlighted with stars surrounded with circles for clear differentiation. Among NASA's confirmed exoplanets, Kepler-158d (radius: 0.43 $R_\oplus$), and Kepler-879c (radius: 0.5 $R_\oplus$) stand as the second and third smallest USPs, across all host star types. Kepler-158d is the smallest USP orbiting K dwarfs. Moreover, Kepler-879c and Kepler-963c (radius: 0.6 $R_\oplus$) are the two smallest USPs orbiting G-type stars.
  } 
  \label{fig:period-radius-distribution}
\end{figure}

Figure \ref{fig:period-radius-distribution} depicts the period distribution for all confirmed USPs in the NASA Confirmed Exoplanet Archive and \textit{Kepler} KOI Catalog, categorized based on host stellar types. The figure demonstrates that our identified USPs predominantly fall into the category of exoplanets with small, sub-Earth radii. This shows the high sensitivity of the GPFC method and its efficacy in detecting small planets which may have eluded earlier searches.

\begin{figure} 
    \centering
    \includegraphics[scale = 0.67]{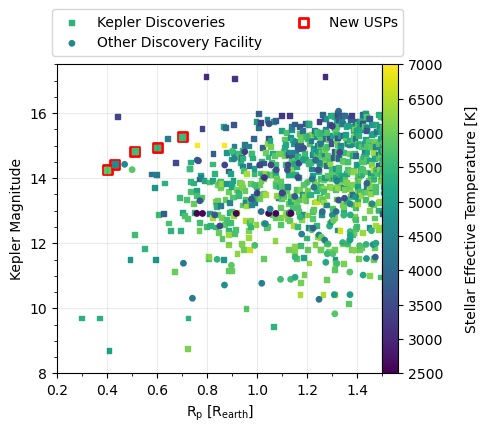}
    \caption{\rt{The planetary radii \rtt{that are $< 1.5$ Earth radii} versus the host star’s \textit{Kepler} magnitude of the five newly discovered USPs (encircled in red) and confirmed exoplanets from the NASA Exoplanet Archive. Marker colors correspond to the host stars' effective temperature. Square markers indicate discoveries in \textit{Kepler} data while circles show discoveries from other facilities. Of the 5,599 confirmed exoplanets in the NASA Exoplanet Archive, 2,276 could not be plotted here due to missing at least one of the three parameters required for this plot.}}
    \label{fig:rp_vs_kepmag_vs_teff}
\end{figure}

Our USPs Kepler-879c with a radius of 0.4 R$_\oplus$, Kepler-158d with a radius of 0.43 R$_\oplus$, and Kepler-1489c with a radius of 0.51 R$_\oplus$ rank as the first, third, and fourth smallest USP planets among NASA's confirmed exoplanets (see Figure \ref{fig:rp_vs_kepmag_vs_teff}). The host star Kepler-158 (KIC 4633570) is a K-type star with an effective stellar temperature of 4896 K, while the host star Kepler-879 (KIC 12266636) is a G-type star with an effective stellar temperature of 5456 K, sourced from the Gaia Archive \citep{gai21}. Furthermore, Kepler-158d stands out as the smallest USP planet orbiting K-type stars.

\begin{figure} 
    \centering
    \includegraphics[scale = 0.67]{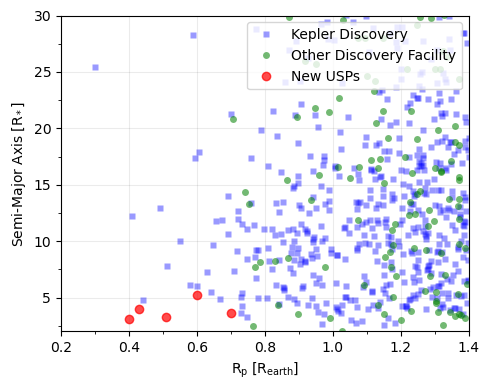}
    \caption{\rt{Orbital semi-major axis normalized by stellar radius (\(a/R_{\star}\)) versus planetary radius (\(R_{p}\)) for a collection of confirmed exoplanets that are $< 1.4$ Earth radii (blue and green markers), with the addition of five newly identified USPs (red dots). Blue squares correspond to discoveries in \textit{Kepler} data while green circles show discoveries from other facilities. Three of the new USPs—Kepler-158d, Kepler-879c, and Kepler-963c—are among the smallest planets that are closest to their host stars, with orbits within 5 stellar radii. Of the 5,599 confirmed exoplanets in the NASA Exoplanet Archive, 1,616 could not be plotted here due to missing at least one of the two parameters required for this plot.}}
    \label{fig:rp_vs_sma_rstar}
\end{figure}

Our USPs Kepler-879c (0.4 R$_\oplus$), Kepler-963c (0.51 R$_\oplus$), KOI-4978.02 (0.6 R$_\oplus$), and Kepler-1489c (0.7R$_\oplus$) are the  four smallest USPs orbiting G-type stars among NASA's confirmed exoplanets (Figure \ref{fig:rp_vs_kepmag_vs_teff}). Kepler-879c, Kepler-158d, Kepler-1489c, and \rtt{KOI-4078.02} are among the smallest planets that are closest to their host stars, with orbits within 5 stellar radii (Figure \ref{fig:rp_vs_sma_rstar}). 

\rt{Table \ref{tab:smallest_5_VK_mag} presents comparisons of Kepler-band magnitudes and spectral types for the stars hosting the current five smallest USPs in the NASA Confirmed Exoplanets Archive. The average radius values were selected from multiple sources listed in the Archive. Notably, three of our newly discovered USPs are among the five smallest confirmed USPs.}

\subsection{Occurrence Rate vs. Stellar Type}

We investigated the correlation between the occurrence rate of USPs and the spectral type of their host stars. According to \citet{Sanchis_Ojeda_2014}, M-dwarfs are approximately 7.3 times more likely than F-dwarfs to host a USP planet, based on \textit{Kepler} data. Further examination of the NASA Confirmed Planet Archive confirmed a decrease in the occurrence rate of USPs as the effective temperature of the host star increases, as outlined in Table \ref{tab:stellar_type_usp_occur_rate}.

Statistics reveal that while 0.7\% of confirmed USPs in the NASA archive orbit F-type stars, a significant 8.0\% revolve around M-type stars, marking them as ten times more prevalent. Our findings introduce an additional USP \rt{in orbit around} a K-type star and four orbiting G-type stars, presenting valuable data for further investigation.

\begin{table}
    \centering
    \caption{USP occurrence rate per stellar spectral type among NASA's confirmed exoplanets.}
    \begin{tabular}{ c | c }
    \hline
    \hline
    \textbf{Stellar Type} & \textbf{USP Occurrence Rate}\\
    \hline
    F & 0.7\%\\
    G & 2.2\%\\
    K & 3.1\%\\
    M & 8.0\%\\
    \hline
    \end{tabular}
    \label{tab:stellar_type_usp_occur_rate}
\end{table}

\subsection{Multiplanet Systems}

Each USP we identified resides in a system already known to harbor at least one other planet. The USPs we discovered are consistently the innermost members of their respective systems and possess the smallest radii.

The NASA Confirmed Exoplanet Archives present that 29.5\% of USPs have at least one sibling planet, and 11.5\% have two or more. In contrast, when examining the entire roster of confirmed USPs in the \textit{Kepler} KOI catalog, the prevalence of multiplanet systems and systems with two or more planets drops to 13.8\% and 5\%, respectively. This disparity could suggest that the \textit{Kepler} survey could potentially underreport multiplanet systems. A detailed breakdown is provided in Table \ref{tab:multiplanet_system_percents}. Our recent discoveries provide valuable data for further studies on multiplanet systems and the theories surrounding their formation.

\begin{table}
    \centering
    \caption{Frequency comparison of confirmed planetary companions in systems with an orbiting USP planet, according to the \textit{Kepler} data set and the NASA Exoplanet Archive.}
    \begin{tabular}{ c | c | c }
    \hline
    \hline
    \textbf{Companions} & \textbf{\textit{Kepler} Systems} & \textbf{NASA Archive Systems} \\
    \hline
    \phantom{$\geq$}0 & 86.2\% & 70.5\% \\
    $\geq$1 & 13.8\% & 29.5\% \\ 
    \phantom{$\geq$}1 & \phantom{1}8.6\% & 18.0\% \\ 
    $\geq$2 & \phantom{1}5.2\% & 11.5\% \\ 
    \hline
    \end{tabular}
    \label{tab:multiplanet_system_percents}
\end{table}

Previous studies have often noted the coexistence of USPs with longer-period planetary companions \citep{Sanchis_Ojeda_2014, Winn_2018, Adams_2021}. As suggested by \citet{Sanchis_Ojeda_2014}, even USP systems lacking known siblings likely harbor additional, non-transiting planets. Assuming that multiplanet interactions—such as tidal migration or dynamical excitation—play a crucial role in positioning USPs onto their current orbits, the detection of a USP would naturally suggest a comprehensive follow-up search for other, yet undetected, sibling planets in the same system. Furthermore, \citet{Winn_2018} emphasized that multiplanet systems containing a USP distinguish themselves from other \textit{Kepler} multiplanet systems. While the typical \textit{Kepler} multiplanet system without a USP exhibits period ratios between adjacent planet pairs ranging from 1.5 to 4, systems with an inner USP typically maintain a larger period ratio, exceeding 3, between the USP and its adjacent outer companion. Our discoveries, outlined in Table \ref{tab:PR_layout_3_planetary_system}, align with these established observations.

\begin{table*}
    \caption{Period and radius layout of the three-planetary systems with an orbiting USP. Currently known three-planetary systems in NASA Confirmed Exoplanets Archive are listed followed up by our new three-planetary systems. The period and radius for each planet are shown to manifest the layout of the three-planetary systems.}
    \begin{tabular}{ | l | c | c | c | c | c | c | c | c | }
        \hline
        \noalign{\vskip 4pt}
        & & & \multicolumn{2}{c|}{\textbf{Planet 1}} & \multicolumn{2}{c|}{\textbf{Planet 2}} & \multicolumn{2}{c|}{\textbf{Planet 3}} \\
        \noalign{\vskip 4pt}
        \multicolumn{1}{c|}{\textbf{Host Star}} & \textbf{KIC} & \textbf{Stellar Type} & \textbf{Period} & \textbf{Radius} & \textbf{Period} & \textbf{Radius} & \textbf{Period} & \textbf{Radius} \\
        & & & \textbf{[days]} & \textbf{[R$_\oplus$]} & \textbf{[days]} & \textbf{[R$_\oplus$]} & \textbf{[days]} & \textbf{[R$_\oplus$]} \\
        \hline
        \hline
        \multicolumn{9}{|c|}{\textbf{NASA Confirmed Exoplanets}} \\
        \hline
        Kepler-42 & 8561063 & M & 0.4530 & 3.19 & \phantom{0}1.214 & 0.87 & \phantom{0}1.87 & 0.69 \\
        YZ Cet & - & M & 0.7520 & - & \phantom{0}1.969 & - & \phantom{0}4.66 & - \\
        TOI-431 & - & K & 0.4900 & 1.28 & \phantom{0}4.849 & - & 12.46 & 3.28 \\
        K2-183 & - & G & 0.4690 & 1.46 & 10.793 & 2.51 & 22.63 & 5.2 \\
        CoRoT-7 & - & G & 0.8540 & 1.58 & \phantom{0}3.700 & - & \phantom{0}8.97 & - \\
        K2-299 & - & G & 0.9116 & 1.35 & \phantom{0}4.508 & 1.7 & 14.64 & 2.94 \\
        \hline
        \hline
        \multicolumn{9}{|c|}{\textbf{New Candidates}} \\
        \hline
        Kepler-158 & 4633570 & K & 0.645088 & 0.43 & 16.709 & 2.27 & 28.55 & 1.97 \\
        Kepler-879 & 12266636 & G & 0.646716 & 0.4 & 12.655 & 1.02 & 33.39 & 2.48 \\
        \hline
    \end{tabular}
    \label{tab:PR_layout_3_planetary_system}
\end{table*}

\section{Conclusion and Future Work}

In this study, we undertook a systematic search for ultra-short period transits within the \textit{Kepler} dataset utilizing our innovative GPFC method. This approach combines the computational efficacy of GPU phase folding with the signal detection capability of a Convolutional Neural Network.
One significant advantage of the GPFC method is its ability to process raw light curves from the \textit{Kepler} survey independently of the TCE catalog. \rtt{Given that the GPFC system is new, we started with a conservative approach to validate it using confirmed and candidate KOIs, as those targets have already undergone vetting by other researchers. The system is capable of completing this search within two days. In our future work, we intend to expand our research to search for transits over extended period ranges and across the larger \textit{Kepler} catalog. We will provide updates on these expansions in our upcoming papers.}

With the GPFC system we report the discovery of five small USPs, namely Kepler-158d, Kepler-963c, Kepler-879c, Kepler-1489c, and KOI-4978.02. Notably, Kepler-879c (0.4 R$_\oplus$), Kepler-158d   (0.43 R$_\oplus$), and Kepler-1489c (0.51 $R_\oplus$), rank as the first, the third and fourth smallest USPs identified to date. Kepler-158d is the smallest USP orbiting K dwarfs, while four of our USPs, Kepler-879c, Kepler-963c, KOI-4978.02, and Kepler-1489c are among the smallest in all known confirmed USPs orbiting G dwarfs in the NASA Confirmed Exoplanet Archive. Additionally, Kepler-879c, Kepler-158d, Kepler-1489c, and Kepler-963c are among the smallest planets that are closest to their host stars, measured in stellar radii.

All five USPs coexist in their respective host systems with other known planets. Notably, both Kepler-158d and Kepler-879c are part of 3-planet multiplanetary systems. These discoveries align with USP formation theories, suggesting that outer planets within the host system may play an important role in positioning the innermost USP planet in a close orbit to the host.

The discovery of the sub-Earth-sized planets demonstrates the effectiveness of the GPFC method for detecting small exoplanets. The GPFC method is generic and can be applied to longer period ranges and many other transit surveys such as TESS (\citet{Ricker-TESS}), and upcoming space transit missions, PLATO (\citet{Rauer-PLATO}) and ET (\citet{GeJ22_Earth2_wp,GeJ22_Earth2_mission, GeJ22_Earth2,GeJ24_ET-space-sci}). We see great potential in its broader applications and we aim to explore these in future studies.

\section*{Acknowledgements}

\rt{We extend our sincere gratitude to the referee for time and expertise in reviewing our manuscript. The insightful feedback provided has been instrumental in improving the clarity and thoroughness of our work. We also wish to thank Prof. Josh Winn for his valuable comments, which have significantly contributed to the overall quality of the paper.} This research has made use of NASA's Astrophysics Data System and the NASA Exoplanet Archive, which is operated by the California Institute of Technology under contract with NASA's Exoplanet Exploration Program. The data presented in this paper includes information collected by the \textit{Kepler} mission. The funding for the \textit{Kepler} mission is provided by NASA's Science Mission Directorate. JG acknowledges the support from the Strategic Priority Program on Space Science of Chinese Academy of Sciences under grant No.XDA15020600. \rt{We made extensive use of the Lightkurve Package \citep{Lightkurve_pkg} as well as other key tools such as NumPy, SciPy, and Matplotlib. Our work would have been much harder without these packages.}

\section*{Data Availability}
The \textit{Kepler} light curves used in this study are available at the following link: \url{https://exoplanetarchive.ipac.caltech.edu}. Additionally, a table listing all confirmed physically periodic sources will be made available online.

\bibliography{refs}
\bibliographystyle{mnras}


\bsp	
\label{lastpage}

\end{document}